# A Modified Method for Deriving Self-Conjugate Dirac Hamiltonians in Arbitrary Gravitational Fields and Its Application to Centrally and Axially Symmetric Gravitational Fields

M.V. Gorbatenko, V.P. Neznamov[1]
Russian Federal Nuclear Center – All-Russian Research Institute of Experimental Physics, 37 Mira Ave., Sarov, 607188 Russia

Abstract

We have proposed previously a method for constructing self-conjugate Hamiltonians $\mathrm{H}_\eta$ in the $\eta$-representation with a flat scalar product to describe the dynamics of Dirac particles in arbitrary gravitational fields. In this paper, we prove that, for block-diagonal metrics, the Hamiltonians $\mathrm{H}_\eta$ can be obtained, in particular, using "reduced" parts of Dirac Hamiltonians, i.e. expressions for Dirac Hamiltonians derived using tetrad vectors in the Schwinger gauge without or with a few summands with bispinor connectivities. Based on these results, we propose a modified method for constructing Hamiltonians in the $\eta$-representation with a significantly smaller amount of required calculations. Using this method, here we for the first time find self-conjugate Hamiltonians for a number of metrics, including the Kerr metric in the Boyer-Lindquist coordinates, the Eddington-Finkelstein, Finkelstein-Lemaitre, Kruskal, Clifford torus metrics and for non-stationary metrics of open and spatially flat Friedmann models.

[1] E-mail: neznamov@vniief.ru

## 1. Introduction

In [1], we proposed a method for constructing self-conjugate Hamiltonians $\mathrm{H}_\eta$ in the $\eta$-representation with a flat scalar product to describe the dynamics of Dirac particles in arbitrary gravitational fields.

Using the algorithm proposed in [1], we calculated Hamiltonians in the $\eta$-representation for the Schwarzschild and Friedmann-Robertson-Walker cosmological model metrics. However, application of the algorithm to the Kerr metric necessitated a large amount of calculations to find Christoffel symbols, bispinor connectivities etc., and cumbersome algebraic transformations of arising expressions.

We made attempts to simplify the algorithm [1]. First, we proved the theorem, according to which a Hamiltonian in the $\eta$-representation for an arbitrary gravitational field, including a time-dependent one, is a Hermitian part of the initial Dirac Hamiltonian $\tilde{H}$ derived using tetrad vectors in the Schwinger gauge[2].

$$\mathrm{H}_\eta = \frac{1}{2}\left(\tilde{H} + \tilde{H}^+\right). \tag{1}$$

Then, for block-diagonal metrics, using Eq. (1), we proved the second theorem, according to which the Hamiltonians $\tilde{H}$ and $\tilde{H}^+$ in Eq. (1) can be replaced by their "reduced" parts without or with a few summands with bispinor connectivities:

$$\mathrm{H}_\eta = \frac{1}{2}\left(\tilde{H}_{red} + \tilde{H}^+_{red}\right) + \Delta\tilde{H}\,. \tag{2}$$

Block-diagonal metrics are understood to be metric tensors of the form

$$g_{\alpha\beta} = \begin{vmatrix} g_{00} & g_{01} & 0 & 0 \\ g_{01} & g_{11} & 0 & 0 \\ 0 & 0 & g_{22} & 0 \\ 0 & 0 & 0 & g_{33} \end{vmatrix}. \tag{3}$$

Apparently, the cases belong to the same kind as (3) when $g_{01} = 0$, and also when $g_{02}$ or $g_{03}$ are used instead of $g_{01}$.

In Eq. (2), $\tilde{H}_{red}$ is part of the initial Dirac Hamiltonian, which contains only the mass term and terms with momentum operator components (i.e. with coordinate derivatives).

The summand $\Delta\tilde{H}$ in (2) equals

$$\Delta\tilde{H} = \frac{i}{4}\left(\frac{\partial\tilde{H}_{\underline{n}0}}{\partial x^p} + \frac{g^{0k}}{g^{00}}\frac{\partial\tilde{H}_{\underline{n}k}}{\partial x^p}\right)\tilde{H}^p_{\underline{m}} S_{\underline{mn}}\,. \tag{4}$$

One can see that $\Delta\tilde{H}$ is a fairly simple expression, which in some cases differs from zero for the block-diagonal metrics with $g_{0k} \neq 0$. For example, the Kerr metric in Boyer-Lindquist coordinates [2], [3] belongs to such case. Of course, application of Eq. (2) makes the procedure of deriving self-conjugate Hamiltonians in the $\eta$-representation much less complicated.

Eqs. (1) and (2) are proven in Sects. 3, 4 of this paper.

In the second part of the paper, we use (1) and (2) to find self-conjugate Hamiltonians $\mathrm{H}_\eta$ for the Kerr [2], [3], Eddington-Finkelstein [4], [5], Painlevé-Gullstrand [6], [7], Finkelstein-Lemaitre [5], Kruskal [8], [9], Clifford torus [10] metrics, and for non-stationary metrics of open and spatially flat Friedmann models. For all these metrics, except for the Painlevé-Gullstrand one, self-conjugate Hamiltonians are derived for the first time[3].

At the end of the paper, we prove that self-conjugate Dirac Hamiltonians in a weak Kerr field are physically equivalent in both harmonic Cartesian and Boyer-Lindquist coordinates.

In the Conclusions, we discuss the outcome of this study and the results of applying the developed algorithm to the evolution of bound atomic and quark states in the expanding universe.

---

[2] We use the same notations as in [1].

[3] For the Painlevé-Gullstrand metric, a physically equivalent self-conjugate Hamiltonian has been derived and studied earlier in [11].

## 2. Reducing the Dirac equation to the Schroedinger form. An algorithm for finding a self-conjugate Hamiltonian in the $\eta$-representation

Let us recall the line of corresponding reasoning and introduce the notation. Tetrad vectors are defined by the relation

$$H_{\underline{\alpha}}^{\mu} H_{\underline{\beta}}^{\nu} g_{\mu\nu} = \eta_{\underline{\alpha\beta}}, \tag{5}$$

where

$$\eta_{\underline{\alpha\beta}} = \mathrm{diag}\left[-1,1,1,1\right]. \tag{6}$$

In addition to the system of tetrad vectors $H_{\underline{\alpha}}^{\mu}$, one can introduce three other systems of tetrad vectors, $H_{\underline{\alpha}\mu}$, $H^{\underline{\alpha}\mu}$, $H_{\mu}^{\underline{\alpha}}$, which differ from $H_{\underline{\alpha}}^{\mu}$ in the location of the global and local (underlined) indices. The global indices are raised up and lowered by means of the metric tensor $g_{\mu\nu}$ and inverse tensor $g^{\mu\nu}$, and the local indices, by means of the tensors $\eta_{\underline{\alpha\beta}}$, $\eta^{\underline{\alpha\beta}}$.

We assume that the quantum mechanical motion of particles is described by the Dirac equation, which is written in the units of $\hbar = c = 1$ as

$$\gamma^{\alpha}\left(\frac{\partial\psi}{\partial x^{\alpha}} + \Phi_{\alpha}\psi\right) - m\psi = 0. \tag{7}$$

Here, $m$ is the particle mass, $\psi$ is a four-component "column" bispinor, and $\gamma^{\alpha}$ are $4\times4$ Dirac matrices, satisfying the relation

$$\gamma^{\alpha}\gamma^{\beta} + \gamma^{\beta}\gamma^{\alpha} = 2g^{\alpha\beta}E. \tag{8}$$

$E$ in (8) means a unity $4\times4$ matrix.

The round parentheses in (7) contain a covariant bispinor derivative, $\nabla_{\alpha}\psi$:

$$\nabla_{\alpha}\psi = \frac{\partial\psi}{\partial x^{\alpha}} + \Phi_{\alpha}\psi. \tag{9}$$

Eq. (9) for $\nabla_{\alpha}\psi$ contains the bispinor connectivity $\Phi_{\alpha}$, for finding which one should fix some system of tetrad vectors $H_{\underline{\alpha}}^{\mu}$ defined as (5). Upon that, the quantity $\Phi_{\alpha}$ can be expressed through "Christoffel" vector derivatives in the following way (the "Christoffel" derivatives are denoted by a semicolon):

$$\Phi_{\alpha} = -\frac{1}{4} H_{\mu}^{\underline{\varepsilon}} H_{\nu\underline{\varepsilon};\alpha} S^{\mu\nu}. \tag{10}$$

The expression for $S^{\mu\nu}$ in (10) is defined below, see (14). The bispinor connectivity $\Phi_{\alpha}$ given by (10) provides invariance of the covariant derivative $\nabla_{\alpha}\psi$ with respect to the transition from one system of tetrad vectors to another.

In what follows, along with Dirac matrices with global indices $\gamma^{\alpha}$, we will use Dirac matrices with local indices $\gamma^{\underline{\alpha}}$. The relation between $\gamma^{\alpha}$ and $\gamma^{\underline{\alpha}}$ is given by the expression

$$\gamma^{\alpha} = H_{\underline{\beta}}^{\alpha}\gamma^{\underline{\beta}}. \tag{11}$$

It follows from (11), (8), (5) that

$$\gamma^{\underline{\alpha}}\gamma^{\underline{\beta}} + \gamma^{\underline{\beta}}\gamma^{\underline{\alpha}} = 2\eta^{\underline{\alpha\beta}}E. \tag{12}$$

In terms of the matrices $\gamma^{\underline{\alpha}}$, the Dirac equation (7) can be written as follows:

$$H_{\underline{\mu}}^{\alpha}\gamma^{\underline{\mu}}\left(\frac{\partial\psi}{\partial x^{\alpha}} + \Phi_{\alpha}\psi\right) - m\psi = 0. \tag{13}$$

It is convenient (but not necessary) to choose the quantities $\gamma^{\underline{\alpha}}$ so that they have the same form for all local frames of reference. Both systems $\gamma^{\underline{\alpha}}$ and $\gamma^{\alpha}$ can be used to construct a full system of $4\times4$ matrices. The full system is, for example, the system

$$E, \quad \gamma_{\alpha}, \quad S_{\alpha\beta} \equiv \frac{1}{2}\left(\gamma_{\alpha}\gamma_{\beta} - \gamma_{\beta}\gamma_{\alpha}\right), \quad \gamma_5 \equiv \gamma_0\gamma_1\gamma_2\gamma_3, \quad \gamma_5\gamma_{\alpha}. \tag{14}$$

Any system of Dirac matrices provides for several discrete automorphisms. We restrict ourselves to the automorphism

$$\gamma_\alpha \to \gamma_\alpha^+ = -D\gamma_\alpha D^{-1}. \qquad (15)$$

The matrix $D$ will be called anti-Hermitizing.

It follows from (7) that the initial Hamiltonian is given by the following expression:

$$H = -\frac{im}{\left(-g^{00}\right)}\gamma^0 + \frac{i}{\left(-g^{00}\right)}\gamma^0\gamma^k\frac{\partial}{\partial x^k} - i\Phi_0 + \frac{i}{\left(-g^{00}\right)}\gamma^0\gamma^k\Phi_k. \qquad (16)$$

The operator $H$ (16) has a meaning of the evolution operator of the wave function of a Dirac particle in the chosen global frame of reference.

Ref. [1] formulates the rules of finding a Hamiltonian in the $\eta$-representation for a Dirac particle in an arbitrary gravitational field. A-priori information, which is assumed to be known, is information about the metric tensor $g_{\alpha\beta}\left(x\right)$, Christoffel symbols $\begin{pmatrix}\lambda \\ \alpha\beta\end{pmatrix}$, local metric tensor $\eta_{\underline{\alpha\beta}}$ and local Dirac matrices $\left\{\gamma_{\underline{\alpha}}\right\}$. These rules are the following:

1) For a gravitational field described by the metric $g_{\alpha\beta}\left(x\right)$, we find a system of tetrads $\left\{\tilde{H}_{\underline{\mu}}^{\alpha}\left(x\right)\right\}$, satisfying the Schwinger gauge. Recall that components of the tetrads $\tilde{H}_{\underline{0}}^{0}, \tilde{H}_{\underline{0}}^{k}$ in this gauge correlate with components of the tensor $g^{\alpha\beta}\left(x\right)$ as follows:

$$\tilde{H}_{\underline{0}}^{0} = \sqrt{-g^{00}};\quad \tilde{H}_{\underline{0}}^{k} = -\frac{g^{0k}}{\sqrt{-g^{00}}}. \qquad (17)$$

$\tilde{H}_{\underline{k}}^{0}$ components are identically zero

$$\tilde{H}_{\underline{k}}^{0} = 0. \qquad (18)$$

In order to find $\tilde{H}_{\underline{m}}^{n}$, we introduce a tensor, $f^{mn}$, with components

$$f^{mn} = g^{mn} - \frac{g^{0m}g^{0n}}{g^{00}}. \qquad (19)$$

The tensor $f^{mn}$ satisfies the condition

$$f^{mn}g_{nk} = \delta_k^m. \qquad (20)$$

As $\tilde{H}_{\underline{m}}^{n}$ we can use any triplet of three-dimensional vectors, satisfying the relation

$$\tilde{H}_{\underline{k}}^{m}\tilde{H}_{\underline{k}}^{n} = f^{mn}. \qquad (21)$$

In what follows, the quantities, dependent on the choice of tetrad vectors, are denoted by a tilde, if they are calculated in the system of tetrad vectors in the Schwinger gauge.

2) In accordance with (16), we write a general expression for the Hamiltonian $\tilde{H}$.

$$\tilde{H} = -\frac{im}{\left(-g^{00}\right)}\tilde{\gamma}^0 + \frac{i}{\left(-g^{00}\right)}\tilde{\gamma}^0\tilde{\gamma}^k\frac{\partial}{\partial x^k} - i\tilde{\Phi}_0 + \frac{i}{\left(-g^{00}\right)}\tilde{\gamma}^0\tilde{\gamma}^k\tilde{\Phi}_k. \qquad (22)$$

Here:

$$\tilde{\gamma}^\alpha = \tilde{H}_{\underline{\beta}}^{\alpha}\gamma^{\underline{\beta}}, \qquad (23)$$

$$\tilde{\Phi}_\alpha = -\frac{1}{4}\tilde{H}_{\mu}^{\varepsilon}\tilde{H}_{\nu\underline{\varepsilon};\alpha}\tilde{S}^{\mu\nu} = \frac{1}{4}\tilde{H}_{\underline{\mu}}^{\varepsilon}\tilde{H}_{\underline{\nu}\varepsilon;\alpha}S^{\underline{\mu\nu}}. \qquad (24)$$

3) The expression for the Hamiltonian $\mathrm{H}_\eta$ equals

$$\mathrm{H}_\eta = \tilde{\eta}\tilde{H}\tilde{\eta}^{-1} + i\tilde{\eta}\frac{\partial\tilde{\eta}^{-1}}{\partial t}, \qquad (25)$$

where the operator $\tilde{\eta}$ is defined by the relation

$$\tilde{\eta} = \left(-g_G\right)^{1/4}\left(-g^{00}\right)^{1/4}. \qquad (26)$$

As distinct from [1], in Eq. (26) we use only the gravitational part of the determinant $g_G$, which is there due to the presence of an external gravitational field. An additional multiplier arises, if we use curvilinear coordinates in accordance with the equality of scalar products for wave functions in the initial and $\eta$-representations [1]:

$$\langle \phi, \psi \rangle_\rho = (\Phi, \Psi). \tag{27}$$

Hence,

$$g_G = \frac{g}{g_c}, \tag{28}$$

where $g_c$ is the determinant, which arises when the volume element is written in curvilinear coordinates. Given that the conditions of coordinate harmonicity [12], [13] are satisfied, $g_c = 1$ for Cartesian coordinates, $g_c = r^2$ for cylindrical coordinates, $g_c = r^4 \sin^2\theta$ is for spherical coordinates etc.

Eqs. (25), (26) define the operator $\mathrm{H}_\eta$, which is the sought Hermitian Hamiltonian in the $\eta$ representation.

Thus,

$$\begin{aligned}\mathrm{H}_\eta = &-\frac{im}{(-g^{00})}\tilde{\gamma}^0 + \frac{i}{(-g^{00})}\tilde{\gamma}^0\tilde{\gamma}^k\frac{\partial}{\partial x^k} - i\tilde{\Phi}_0 + \frac{i}{(-g^{00})}\tilde{\gamma}^0\tilde{\gamma}^k\tilde{\Phi}_k \\ &-\frac{i}{4(-g^{00})}\tilde{\gamma}^0\tilde{\gamma}^k\frac{\partial \ln(g_G g^{00})}{\partial x^k} + \frac{i}{4}\frac{\partial \ln(g_G g^{00})}{\partial t}.\end{aligned} \tag{29}$$

Note that procedures for constructing self-conjugate Hamiltonians with a flat scalar product that could be used for studying the dynamics of spin ½ particles in gravitational fields of particular form have been proposed in literature more than once [14] - [17]. These attempts are not general, but they produce correct results as applied to the choice of particular metrics and tetrad vectors.

## 3. Proving the equality $\mathrm{H}_\eta = \frac{1}{2}\left(\tilde{H} + \tilde{H}^+\right)$ for an arbitrary, including time-dependent, gravitational field

We start the proof from transforming the right side in (25).

$$\mathrm{H}_\eta = \tilde{\eta}\tilde{H}\tilde{\eta}^{-1} + i\frac{\partial \tilde{\eta}}{\partial t}\tilde{\eta}^{-1}. \tag{30}$$

After inserting (22), (26) into (25) we have:

$$\mathrm{H}_\eta = \tilde{H} - \frac{i}{(-g^{00})}\left(-g^{00}\right)^{1/2}\gamma^{\underline{0}}\tilde{\gamma}^k\frac{1}{4}\frac{\partial \ln(g^{00}g_G)}{\partial x^k} + \frac{i}{4}\frac{\partial \ln(g^{00}g_G)}{\partial t}. \tag{31}$$

The next step in the proof is to find an expression for $\tilde{H}^+$. For this purpose, we employ relations (77) from [1]:

$$\tilde{H}^+ = \tilde{\rho}\tilde{H}\tilde{\rho}^{-1} + \tilde{\Delta}, \tag{32}$$

$$\tilde{\rho} = \left(g^{00}g_G\right)^{1/2}, \tag{33}$$

$$\tilde{\Delta} = \frac{i}{2}\frac{\partial \ln(g^{00}g_G)}{\partial t}. \tag{34}$$

We insert (33), (34) into (32):

$$\tilde{H}^+ = \tilde{H} - \frac{i}{(-g^{00})}\left(-g^{00}\right)^{1/2}\gamma^{\underline{0}}\tilde{\gamma}^k\frac{1}{2}\frac{\partial \ln(g^{00}g_G)}{\partial x^k} + \frac{i}{2}\frac{\partial \ln(g^{00}g_G)}{\partial t}. \tag{35}$$

Using (22) and (35), we calculate the quantity $\frac{1}{2}\left(\tilde{H} + \tilde{H}^+\right)$:

$$\frac{1}{2}\left(\tilde{H}+\tilde{H}^{+}\right)=\tilde{H}-\frac{i}{\left(-g^{00}\right)}\left(-g^{00}\right)^{1/2}\gamma^{\underline{0}}\tilde{\gamma}^{k}\frac{1}{4}\frac{\partial\ln\left(g^{00}g_{G}\right)}{\partial x^{k}}+\frac{i}{4}\frac{\partial\ln\left(g^{00}g_{G}\right)}{\partial t}. \quad (36)$$

By comparing (36) with (31) we conclude that Eq. (1) is valid.

$$\mathrm{H}_{\eta}=\frac{1}{2}\left(\tilde{H}+\tilde{H}^{+}\right). \quad (37)$$

**4. Proving the equality $\mathrm{H}_{\eta}=\frac{1}{2}\left(\tilde{H}_{red}+\tilde{H}_{red}^{+}\right)+\Delta\tilde{H}$ for gravitational fields with the block-diagonal metrics**

An expression for the "reduced" Hamiltonian $\tilde{H}_{red}$ is derived from (22) by deleting the terms with bispinor connectivities. Thus,

$$\tilde{H}_{red}=-\frac{im}{\left(-g^{00}\right)}\tilde{\gamma}^{0}+\frac{i}{\left(-g^{00}\right)}\tilde{\gamma}^{0}\tilde{\gamma}^{k}\frac{\partial}{\partial x^{k}}. \quad (38)$$

This expression can also be written as

$$\tilde{H}_{red}=\tilde{H}+i\tilde{\Phi}_{0}-\frac{i}{\left(-g^{00}\right)}\tilde{\gamma}^{0}\tilde{\gamma}^{k}\tilde{\Phi}_{k}. \quad (39)$$

Taking the Hermitian conjugation from (39), we obtain

$$\tilde{H}_{red}^{+}=\tilde{H}^{+}-i\gamma_{\underline{0}}\tilde{\Phi}_{0}\gamma_{\underline{0}}+\frac{i}{\left(-g^{00}\right)^{1/2}}\gamma_{\underline{0}}\tilde{\Phi}_{k}\tilde{\gamma}^{k}. \quad (40)$$

It follows from (39), (40) that

$$\left(\tilde{H}_{red}+\tilde{H}_{red}^{+}\right)=\left(\tilde{H}+\tilde{H}^{+}\right)-i\gamma_{\underline{0}}\left[\gamma_{\underline{0}},\tilde{\Phi}_{0}\right]_{+}+\frac{i}{\left(-g^{00}\right)^{1/2}}\gamma_{\underline{0}}\left[\tilde{\gamma}^{k},\tilde{\Phi}_{k}\right]_{+}. \quad (41)$$

Considering Eq. (37), we obtain

$$\mathrm{H}_{\eta}=\frac{1}{2}\left(\tilde{H}_{red}+\tilde{H}_{red}^{+}\right)+\frac{i}{2}\gamma_{\underline{0}}\left[\gamma_{\underline{0}},\tilde{\Phi}_{0}\right]_{+}-\frac{i}{2\left(-g^{00}\right)^{1/2}}\gamma_{\underline{0}}\left[\tilde{\gamma}^{k},\tilde{\Phi}_{k}\right]_{+}. \quad (42)$$

Let us introduce the following notation:

$$Y\equiv\frac{i}{2}\gamma_{\underline{0}}\left[\gamma_{\underline{0}},\tilde{\Phi}_{0}\right]_{+}, \quad (43)$$

$$Z\equiv-\frac{i}{2\left(-g^{00}\right)^{1/2}}\gamma_{\underline{0}}\left[\tilde{\gamma}^{k},\tilde{\Phi}_{k}\right]_{+}. \quad (44)$$

Some transformations give the following expressions for $Y$

$$Y=-\frac{i}{4}\tilde{H}_{\underline{m}}^{p}\tilde{H}_{\underline{n}p;0}S_{\underline{mn}}, \quad (45)$$

and for $Z$ :

$$Z=Z_{1}+Z_{2}+Z_{3}+Z_{4}, \quad (46)$$

where

$$Z_{1}=-\frac{i}{4\left(-g^{00}\right)^{1/2}}\tilde{H}_{\underline{0}}^{k}\tilde{H}_{\underline{m}}^{\varepsilon}\tilde{H}_{\underline{n}\varepsilon;k}S_{\underline{mn}}, \quad (47)$$

$$Z_{2}=\frac{i}{4\left(-g^{00}\right)^{1/2}}\tilde{H}_{\underline{m}}^{p}\tilde{H}_{\underline{0}}^{\varepsilon}\tilde{H}_{\underline{n}\varepsilon;p}S_{\underline{mn}}, \quad (48)$$

$$Z_{3}=-\frac{i}{4\left(-g^{00}\right)^{1/2}}\tilde{H}_{\underline{m}}^{p}\tilde{H}_{\underline{n}}^{\varepsilon}\tilde{H}_{\underline{0}\varepsilon;p}S_{\underline{mn}}, \quad (49)$$

$$Z_4 = \frac{i}{4\left(-g^{00}\right)^{1/2}} \tilde{H}^{k}_{\underline{m}} \tilde{H}^{\varepsilon}_{\underline{p}} \tilde{H}_{\underline{q}\varepsilon;k} \varepsilon_{\underline{mpq}} \gamma_{\underline{5}} . \tag{50}$$

In (50), $\gamma_{\underline{5}} = \gamma_{\underline{0}}\gamma_{\underline{1}}\gamma_{\underline{2}}\gamma_{\underline{3}}$, $\varepsilon_{\underline{mpq}}$ is a totally antisymmetric third-rank tensor.

Then, we calculate $Y$ and $Z$ using the relations (3), (5), (17), (18), and diagonal representation of $\left\{H^{m}_{\underline{k}}\right\}$. Direct calculations show that

$$\Delta\tilde{H} = Y + Z_1 + Z_2 = \frac{i}{4}\left[\frac{\partial \tilde{H}_{\underline{n}0}}{\partial x^p} + \frac{g^{0k}}{g^{00}}\frac{\partial \tilde{H}_{\underline{n}k}}{\partial x^p}\right]\tilde{H}^{p}_{\underline{m}} S_{\underline{mn}}, \tag{51}$$

$$Z_3 = Z_4 = 0 . \tag{52}$$

Thus, it turns out that for the block-diagonal metrics of the form (3) we can find the Hamiltonian $\mathrm{H}_\eta$ using the fairly simple formula (2).

### 5. Centrally symmetric gravitational field

This section presents Hamiltonians in the $\eta$-representation for Dirac particles in centrally symmetric gravitational fields, when the metrics are written in various coordinates.

#### 5.1. The Schwarzschild metric

Writing the Schwarzschild solution in the coordinates

$$\left(x^0, x^1, x^2, x^3\right) \equiv \left(t, r, \theta, \varphi\right) \tag{53}$$

gives:

$$ds^2 = -\left(1-\frac{r_0}{r}\right)dt^2 + \frac{dr^2}{\left(1-\frac{r_0}{r}\right)} + r^2\left(d\theta^2 + \sin^2\theta\, d\varphi^2\right). \tag{54}$$

In Eq. (54), $r_0$ is the gravitational radius $\left(r_0 = 2M\right)$.

The resulting expression for $\mathrm{H}_\eta$ derived in [1] and revised to include (26) is

$$\mathrm{H}_\eta = im\sqrt{f}\gamma_{\underline{0}} - i\sqrt{f}\gamma_{\underline{0}}\left\{\gamma_{\underline{1}}\sqrt{f}\left(\frac{\partial}{\partial r}+\frac{1}{r}\right) + \gamma_{\underline{2}}\frac{1}{r}\left(\frac{\partial}{\partial\theta}+\frac{1}{2}\mathrm{ctg}\theta\right) + \gamma_{\underline{3}}\frac{1}{r\cdot\sin\theta}\frac{\partial}{\partial\varphi}\right\} - \frac{i}{2}\frac{\partial f}{\partial r}\cdot\gamma_{\underline{0}}\gamma_{\underline{1}} . \tag{55}$$

In Eq. (55), $f = 1 - \frac{r_0}{r}$.

It is easy to verify that Eq. (55) can be found in a comparatively straightforward manner using formula (2), if we take into account that

$$\tilde{H}_{red} = im\sqrt{f}\gamma_{\underline{0}} - i\sqrt{f}\gamma_{\underline{0}}\left\{\gamma_{\underline{1}}\sqrt{f}\frac{\partial}{\partial r} + \gamma_{\underline{2}}\frac{1}{r}\frac{\partial}{\partial\theta} + \gamma_{\underline{3}}\frac{1}{r\cdot\sin\theta}\frac{\partial}{\partial\varphi}\right\},\ \Delta\tilde{H} = 0 . \tag{56}$$

In Refs. [14], [1], the authors also derived a Hamiltonian for the Schwarzschield metric in isotropic coordinates

$$ds^2 = -\frac{\left(1-\frac{r_0}{4R}\right)^2}{\left(1+\frac{r_0}{4R}\right)^2}dt^2 + \left(1+\frac{r_0}{4R}\right)^4\left(dx^2+dy^2+dz^2\right), \tag{57}$$

$$\mathrm{H}_\eta = im\frac{\left(1-\frac{r_0}{4R}\right)}{\left(1+\frac{r_0}{4R}\right)}\gamma_{\underline{0}} - i\frac{\left(1-\frac{r_0}{4R}\right)}{\left(1+\frac{r_0}{4R}\right)^3}\gamma_{\underline{0}}\gamma^{\underline{k}}\frac{\partial}{\partial x^k} - i\frac{\left(1-\frac{r_0}{8R}\right)}{\left(1+\frac{r_0}{4R}\right)^4}\frac{r_0}{2}\frac{R_k}{R^3}\gamma_{\underline{0}}\gamma^{\underline{k}} . \tag{58}$$

The expression for $\mathrm{H}_\eta$ can be easily derived from (2) using

$$\tilde{\mathrm{H}}_{red} = im\frac{\left(1-\frac{r_0}{4R}\right)}{\left(1+\frac{r_0}{4R}\right)}\gamma_{\underline{0}} - i\frac{\left(1-\frac{r_0}{4R}\right)}{\left(1+\frac{r_0}{4R}\right)^3}\gamma_{\underline{0}}\gamma^{\underline{k}}\frac{\partial}{\partial x^k},\ \Delta\tilde{H} = 0\,. \tag{59}$$

### 5.2. Eddington-Finkelstein metric

The Eddington-Finkelstein solution ([4], [5]) in the coordinates

$$\left(x^0, x^1, x^2, x^3\right) \equiv \left(t, r, \theta, \varphi\right) \tag{60}$$

is given by

$$g_{\alpha\beta} = \begin{pmatrix} -\left(1-\frac{r_0}{r}\right) & \frac{r_0}{r} & 0 & 0 \\ \frac{r_0}{r} & \left(1+\frac{r_0}{r}\right) & 0 & 0 \\ 0 & 0 & r^2 & 0 \\ 0 & 0 & 0 & r^2\sin^2\theta \end{pmatrix}. \tag{61}$$

$$g = -r^4\cdot\sin^2\theta,\ g_G = -1\,. \tag{62}$$

The inverse tensor has the following form:

$$g^{\alpha\beta} = \begin{pmatrix} -\left(1+\frac{r_0}{r}\right) & \frac{r_0}{r} & 0 & 0 \\ \frac{r_0}{r} & \left(1-\frac{r_0}{r}\right) & 0 & 0 \\ 0 & 0 & \frac{1}{r^2} & 0 \\ 0 & 0 & 0 & \frac{1}{r^2\sin^2\theta} \end{pmatrix}. \tag{63}$$

Table 1. $\tilde{H}^{\alpha}_{\underline{\mu}}$ -like tetrad vectors

| Tetrad vectors | Tetrad vector components | | | |
|---|---|---|---|---|
| $\tilde{H}^{\alpha}_{\underline{0}}$ | $\tilde{H}^{0}_{\underline{0}} = \sqrt{\left(1+\frac{r_0}{r}\right)}$ | $\tilde{H}^{1}_{\underline{0}} = -\frac{r_0}{r\sqrt{\left(1+\frac{r_0}{r}\right)}}$ | $\tilde{H}^{2}_{\underline{0}} = 0$ | $\tilde{H}^{3}_{\underline{0}} = 0$ |
| $\tilde{H}^{\alpha}_{\underline{1}}$ | $\tilde{H}^{0}_{\underline{1}} = 0$ | $\tilde{H}^{1}_{\underline{1}} = \frac{1}{\sqrt{\left(1+\frac{r_0}{r}\right)}}$ | $\tilde{H}^{2}_{\underline{1}} = 0$ | $\tilde{H}^{3}_{\underline{1}} = 0$ |
| $\tilde{H}^{\alpha}_{\underline{2}}$ | $\tilde{H}^{0}_{\underline{2}} = 0$ | $\tilde{H}^{1}_{\underline{2}} = 0$ | $\tilde{H}^{2}_{\underline{2}} = \frac{1}{r}$ | $\tilde{H}^{3}_{\underline{2}} = 0$ |
| $\tilde{H}^{\alpha}_{\underline{3}}$ | $\tilde{H}^{0}_{\underline{3}} = 0$ | $\tilde{H}^{1}_{\underline{3}} = 0$ | $\tilde{H}^{2}_{\underline{3}} = 0$ | $\tilde{H}^{3}_{\underline{3}} = \frac{1}{r\sin\theta}$ |

Calculations of a "reduced" Hamiltonian using (38) gives

$$\tilde{\mathrm{H}}_{red} = \frac{im}{\sqrt{\left(1+\frac{r_0}{r}\right)}}\gamma_{\underline{0}} + \frac{ir_0}{r\left(1+\frac{r_0}{r}\right)}\frac{\partial}{\partial r} - \frac{i}{\sqrt{\left(1+\frac{r_0}{r}\right)}}\gamma_{\underline{0}}\left(\frac{1}{\sqrt{\left(1+\frac{r_0}{r}\right)}}\gamma_{\underline{1}}\frac{\partial}{\partial r} + \gamma_{\underline{2}}\frac{1}{r}\frac{\partial}{\partial\theta} + \gamma_{\underline{3}}\frac{1}{r\sin\theta}\frac{\partial}{\partial\varphi}\right). \tag{64}$$

The Hamiltonian in the $\eta$-representation is calculated using (2) given that $\Delta\tilde{H} = 0$ for the metric of interest. We obtain:

$$H_\eta = i\gamma_{\underline{0}}\frac{m}{\sqrt{\left(1+\frac{r_0}{r}\right)}} - i\gamma_{\underline{0}}\gamma_{\underline{1}}\frac{1}{\left(1+\frac{r_0}{r}\right)}\left(\left(\frac{\partial}{\partial r}+\frac{1}{r}\right)+\frac{r_0}{2r^2}\frac{1}{\left(1+\frac{r_0}{r}\right)}\right) - $$
$$-i\gamma_{\underline{0}}\gamma_{\underline{2}}\frac{1}{\sqrt{\left(1+\frac{r_0}{r}\right)}}\frac{1}{r}\left(\frac{\partial}{\partial\theta}+\frac{1}{2}\mathrm{ctg}\theta\right) - i\gamma_{\underline{0}}\gamma_{\underline{3}}\frac{1}{\sqrt{\left(1+\frac{r_0}{r}\right)}}\frac{1}{r\sin\theta}\frac{\partial}{\partial\varphi} + \tag{65}$$
$$+i\frac{r_0}{r}\frac{1}{\left(1+\frac{r_0}{r}\right)}\left(\left(\frac{\partial}{\partial r}+\frac{1}{r}\right) - \frac{1}{2r\left(1+\frac{r_0}{r}\right)}\right).$$

### 5.3. Painlevé-Gullstrand metric

In this section, we find a self-conjugate Hamiltonian $\mathrm{H}_\eta$ for a Dirac particle in a spherically symmetric gravitational field described by the Painlevé-Gullstrand metric. The Hamiltonian $\mathrm{H}_\eta$ for this metric is calculated first using the algorithm of [1] and then using (1) and (2).

In the $(t, r, \theta, \varphi)$ coordinates, the Painlevé-Gullstrand metric [6] has the following form:

$$g_{\alpha\beta} = \begin{pmatrix} -\left(1-\frac{r_0}{r}\right) & \sqrt{\frac{r_0}{r}} & 0 & 0 \\ \sqrt{\frac{r_0}{r}} & 1 & 0 & 0 \\ 0 & 0 & r^2 & 0 \\ 0 & 0 & 0 & r^2\sin^2\theta \end{pmatrix} \quad g^{\alpha\beta} = \begin{pmatrix} -1 & \sqrt{\frac{r_0}{r}} & 0 & 0 \\ \sqrt{\frac{r_0}{r}} & \left(1-\frac{r_0}{r}\right) & 0 & 0 \\ 0 & 0 & \frac{1}{r^2} & 0 \\ 0 & 0 & 0 & \frac{1}{r^2\sin^2\theta} \end{pmatrix}. \tag{66}$$

The determinants equal

$$g = -r^4\sin^2\theta, \quad g_G = -1. \tag{67}$$

Tetrad vectors in the Schwinger gauge:

$$\left.\begin{array}{l} \tilde{H}^0_{\underline{0}} = 1, \quad \tilde{H}^1_{\underline{0}} = -\sqrt{\dfrac{r_0}{r}}, \quad \tilde{H}^2_{\underline{0}} = 0, \quad \tilde{H}^3_{\underline{0}} = 0, \\ \tilde{H}^0_{\underline{1}} = 0, \quad \tilde{H}^1_{\underline{1}} = 1, \quad \tilde{H}^2_{\underline{1}} = 0, \quad \tilde{H}^3_{\underline{1}} = 0, \\ \tilde{H}^0_{\underline{2}} = 0, \quad \tilde{H}^1_{\underline{2}} = 0, \quad \tilde{H}^2_{\underline{2}} = \dfrac{1}{r}, \quad \tilde{H}^3_{\underline{2}} = 0, \\ \tilde{H}^0_{\underline{3}} = 0, \quad \tilde{H}^1_{\underline{3}} = 0, \quad \tilde{H}^2_{\underline{3}} = 0, \quad \tilde{H}^3_{\underline{3}} = \dfrac{1}{r \cdot \sin\theta} \end{array}\right\}. \tag{68}$$

$$\left.\begin{array}{l} \tilde{H}_{\underline{0}0} = -1, \quad \tilde{H}_{\underline{0}1} = 0, \quad \tilde{H}_{\underline{0}2} = 0, \quad \tilde{H}_{\underline{0}3} = 0, \\ \tilde{H}_{\underline{1}0} = \sqrt{\dfrac{r_0}{r}}, \quad \tilde{H}_{\underline{1}1} = 1, \quad \tilde{H}_{\underline{1}2} = 0, \quad \tilde{H}_{\underline{1}3} = 0, \\ \tilde{H}_{\underline{2}0} = 0, \quad \tilde{H}_{\underline{2}1} = 0, \quad \tilde{H}_{\underline{2}2} = r, \quad \tilde{H}_{\underline{2}3} = 0, \\ \tilde{H}_{\underline{3}0} = 0, \quad \tilde{H}_{\underline{3}1} = 0, \quad \tilde{H}_{\underline{3}2} = 0, \quad \tilde{H}_{\underline{3}3} = r \cdot \sin\theta \end{array}\right\}. \tag{69}$$

Christoffel symbols:

$$\left.\begin{array}{l} \begin{pmatrix} 0 \\ 00 \end{pmatrix} = \dfrac{1}{2r^2}\sqrt{\dfrac{r_0^3}{r}} \\ \begin{pmatrix} 0 \\ 01 \end{pmatrix} = \dfrac{1}{2}\dfrac{r_0}{r^2} \\ \begin{pmatrix} 0 \\ 11 \end{pmatrix} = \dfrac{1}{2r}\sqrt{\dfrac{r_0}{r}} \\ \begin{pmatrix} 0 \\ 02 \end{pmatrix} = \begin{pmatrix} 0 \\ 03 \end{pmatrix} = \begin{pmatrix} 0 \\ 12 \end{pmatrix} = \begin{pmatrix} 0 \\ 13 \end{pmatrix} = \begin{pmatrix} 0 \\ 23 \end{pmatrix} = 0 \\ \begin{pmatrix} 0 \\ 22 \end{pmatrix} = -\sqrt{r_0 r} \\ \begin{pmatrix} 0 \\ 33 \end{pmatrix} = -\sqrt{r_0 r}\sin^2\theta \end{array}\right\}. \tag{70}$$

$$\left.\begin{array}{l} \begin{pmatrix} 1 \\ 00 \end{pmatrix} = \dfrac{1}{2}\dfrac{r_0}{r^2}\left(1 - \dfrac{r_0}{r}\right) \\ \begin{pmatrix} 1 \\ 01 \end{pmatrix} = -\dfrac{1}{2}\dfrac{r_0^{3/2}}{r^{5/2}} \\ \begin{pmatrix} 1 \\ 11 \end{pmatrix} = -\dfrac{1}{2}\dfrac{r_0}{r^2} \\ \begin{pmatrix} 1 \\ 02 \end{pmatrix} = \begin{pmatrix} 1 \\ 03 \end{pmatrix} = \begin{pmatrix} 1 \\ 12 \end{pmatrix} = \begin{pmatrix} 1 \\ 13 \end{pmatrix} = \begin{pmatrix} 1 \\ 23 \end{pmatrix} = 0 \\ \begin{pmatrix} 1 \\ 22 \end{pmatrix} = -r\left(1 - \dfrac{r_0}{r}\right) \\ \begin{pmatrix} 1 \\ 33 \end{pmatrix} = -r\sin^2\theta\left(1 - \dfrac{r_0}{r}\right) \end{array}\right\}. \tag{71}$$

$$\left.\begin{array}{l} \begin{pmatrix} 2 \\ 12 \end{pmatrix} = \frac{1}{r} \\ \begin{pmatrix} 2 \\ 00 \end{pmatrix} = \begin{pmatrix} 2 \\ 01 \end{pmatrix} = \begin{pmatrix} 2 \\ 11 \end{pmatrix} = \begin{pmatrix} 2 \\ 02 \end{pmatrix} = \begin{pmatrix} 2 \\ 03 \end{pmatrix} = \begin{pmatrix} 2 \\ 13 \end{pmatrix} = \begin{pmatrix} 2 \\ 22 \end{pmatrix} = \begin{pmatrix} 2 \\ 23 \end{pmatrix} = 0 \\ \begin{pmatrix} 2 \\ 33 \end{pmatrix} = -\sin\theta\cos\theta. \end{array}\right\}. \qquad (72)$$

$$\left.\begin{array}{l} \begin{pmatrix} 3 \\ 13 \end{pmatrix} = \frac{1}{r} \\ \begin{pmatrix} 3 \\ 00 \end{pmatrix} = \begin{pmatrix} 3 \\ 01 \end{pmatrix} = \begin{pmatrix} 3 \\ 11 \end{pmatrix} = \begin{pmatrix} 3 \\ 02 \end{pmatrix} = \begin{pmatrix} 3 \\ 12 \end{pmatrix} = \begin{pmatrix} 3 \\ 03 \end{pmatrix} = \begin{pmatrix} 3 \\ 22 \end{pmatrix} = \begin{pmatrix} 3 \\ 33 \end{pmatrix} = 0 \\ \begin{pmatrix} 3 \\ 23 \end{pmatrix} = \frac{\cos\theta}{\sin\theta} \end{array}\right\}. \qquad (73)$$

Bispinor connectivities are calculated by the formula (24) using (66) - (73). We obtain:

$$\left.\begin{array}{l} \tilde{\Phi}_0 = \frac{1}{4}\frac{r_0}{r^2} S_{\underline{0}\underline{1}} \\ \tilde{\Phi}_1 = \frac{1}{4}\frac{1}{r}\sqrt{\frac{r_0}{r}} S_{\underline{0}\underline{1}} \\ \tilde{\Phi}_2 = -\frac{1}{2}\sqrt{\frac{r_0}{r}} S_{\underline{0}\underline{2}} - \frac{1}{2} S_{\underline{1}\underline{2}} \\ \tilde{\Phi}_3 = -\frac{1}{2}\sqrt{\frac{r_0}{r}} \sin\theta \cdot S_{\underline{0}\underline{3}} + \frac{1}{2}\sin\theta \cdot S_{\underline{3}\underline{1}} - \frac{1}{2}\cos\theta \cdot S_{\underline{2}\underline{3}} \end{array}\right\}. \qquad (74)$$

In order to find the Hamiltonian in the $\eta$-representation, $\left(-g^{00}\right) = 1$, expressions for $\tilde{\gamma}^{\alpha} = \tilde{H}^{\alpha}_{\underline{\mu}} \gamma^{\underline{\mu}}$ and expressions (74) for $\tilde{\Phi}_{\alpha}$ are put into the primary Dirac Hamiltonian $\tilde{H}$,

$$\tilde{H} = -\frac{im}{\left(-g^{00}\right)}\tilde{\gamma}^0 + \frac{i}{\left(-g^{00}\right)}\tilde{\gamma}^0\tilde{\gamma}^k \frac{\partial}{\partial x^k} - i\tilde{\Phi}_0 + \frac{i}{\left(-g^{00}\right)}\tilde{\gamma}^0\tilde{\gamma}^k\tilde{\Phi}_k. \qquad (75)$$

These transformations give

$$\tilde{H} = im\gamma_{\underline{0}} - i\gamma_{\underline{0}}\left\{\gamma_{\underline{1}}\left(\frac{\partial}{\partial r} + \frac{1}{r}\right) + \frac{1}{r}\gamma_{\underline{2}}\left(\frac{\partial}{\partial\theta} + \frac{1}{2}\mathrm{ctg}\theta\right) + \frac{1}{r\sin\theta}\gamma_{\underline{3}}\frac{\partial}{\partial\varphi}\right\} + i\sqrt{\frac{r_0}{r}}\frac{\partial}{\partial r} + i\frac{3}{4}\frac{1}{r}\sqrt{\frac{r_0}{r}}. \qquad (76)$$

The operator $\tilde{\eta}$ for the Painlevé-Gullstrand metric equals

$$\tilde{\eta} = \left(g_G g^{00}\right)^{1/2} = 1, \qquad (77)$$

so the $\eta$-representation coincides with the representation of the Hamiltonian $\tilde{H}$.

The Hamiltonian $\tilde{H}$ (76) is self-conjugate. It is evident that the formula (1) is valid in this case.

It is easy to obtain (76) from (2) given that $\Delta\tilde{H} = 0$, and $\tilde{H}_{red}$ is written as

$$\tilde{H}_{red} = im\gamma_{\underline{0}} - i\gamma_{\underline{0}}\left\{\gamma_{\underline{1}}\frac{\partial}{\partial r} + \gamma_{\underline{2}}\frac{1}{r}\frac{\partial}{\partial\theta} + \gamma_{\underline{3}}\frac{1}{r\cdot\sin\theta}\frac{\partial}{\partial\varphi}\right\} + i\sqrt{\frac{r_0}{r}}\frac{\partial}{\partial r}. \qquad (78)$$

Thus, as applied to the Painlevé-Gullstrand metric, the same Hamiltonian $H_{\eta}$ was obtained both by the standard algorithm and in a simpler manner using (2).

In [11], a self-conjugate Hamiltonian was obtained for the Painlevé-Gullstrand metric using tetrad vectors in the Schwinger gauge with a set of local Dirac matrices written in spherical coordinates.

$$\left.\begin{aligned}
&\gamma_{\underline{0}} = \gamma_{\underline{0}} \\
&\gamma_{\underline{r}} = \sin\theta\left[\gamma_{\underline{1}}\cos\varphi + \gamma_{\underline{2}}\sin\varphi\right] + \gamma_{\underline{3}}\cos\theta = R\gamma_{\underline{1}}R^{-1} \\
&\gamma_{\underline{\theta}} = \cos\theta\left[\gamma_{\underline{1}}\cos\varphi + \gamma_{\underline{2}}\sin\varphi\right] - \gamma_{\underline{3}}\sin\theta = R\gamma_{\underline{2}}R^{-1} \\
&\gamma_{\underline{\varphi}} = -\gamma_{\underline{1}}\sin\varphi + \gamma_{\underline{2}}\cos\varphi = R\gamma_{\underline{3}}R^{-1}
\end{aligned}\right\}. \tag{79}$$

The set $\left\{\gamma_{\underline{r}}, \gamma_{\underline{\theta}}, \gamma_{\underline{\varphi}}\right\}$ is related to the set $\left\{\gamma_{\underline{1}}, \gamma_{\underline{2}}, \gamma_{\underline{3}}\right\}$ through a unitary matrix $R$,

$$\begin{aligned}
&R = R_1 T_1 R_2 T_2, \\
&R_1 = \exp\left(-\frac{\varphi}{2}\gamma_{\underline{1}}\gamma_{\underline{2}}\right);\ T_1 = \frac{1}{\sqrt{2}}\gamma_{\underline{5}}\gamma_{\underline{1}}\left(E + \gamma_{\underline{1}}\gamma_{\underline{2}}\right), \\
&R_2 = \exp\left(-\frac{\theta}{2}\gamma_{\underline{2}}\gamma_{\underline{3}}\right);\ T_2 = \frac{1}{\sqrt{2}}\gamma_{\underline{5}}\gamma_{\underline{2}}\left(E + \gamma_{\underline{3}}\gamma_{\underline{1}}\right).
\end{aligned} \tag{80}$$

The Hamiltonian from [11] can be written as

$$\tilde{H}_D = im\gamma_{\underline{0}} - i\gamma_{\underline{0}}\left\{\gamma_{\underline{r}}\frac{\partial}{\partial r} + \gamma_{\underline{\theta}}\frac{1}{r}\frac{\partial}{\partial\theta} + \gamma_{\underline{\varphi}}\frac{1}{r\sin\theta}\frac{\partial}{\partial\varphi}\right\} + i\sqrt{\frac{r_0}{r}}\left(\frac{\partial}{\partial r} + \frac{3}{4r}\right). \tag{81}$$

The Hamiltonians (76) and (81) are physically equivalent, because they are related through a unitary transformation,

$$\tilde{H}_D = R\tilde{H}R^{-1}, R^{+} = R^{-1}. \tag{82}$$

Generally speaking, all Hamiltonians in the Schwinger gauge are connected with each other by physically equivalent matrices of spatial rotation. This is what we meant [1] speaking about the uniqueness of Hamiltonians in the $\eta$-representation (see the comments by M.Arminjon in [18]).

### 5.4. Finkelstein-Lemaitre metric

It is of independent interest to study the motion of a Dirac particle in the nonstationary Finkelstein-Lemaitre metric [5], because the time coordinate in this metric coincides with the proper time.

$$ds^2 = -dt^2 + \frac{dr^2}{\left[\frac{3}{2r_0}(r-t)\right]^{2/3}} + \left[\frac{3}{2}(r-t)\right]^{4/3} r_0^{2/3}\left(d\theta^2 + \sin^2\theta\cdot d\varphi^2\right). \tag{83}$$

The determinants equal

$$\left.\begin{aligned}
g &= -\left[\frac{3}{2}(r-t)\right]^2 r_0^2\sin^2\theta \\
g_G &= -\left[\frac{3}{2}(r-t)\right]^2 \frac{r_0^2}{r^4}
\end{aligned}\right\}. \tag{84}$$

Non-zero components of tetrad vectors in the Schwinger gauge:

$$\tilde{H}^0_{\underline{0}} = 1,\quad \tilde{H}^1_{\underline{1}} = \left[\frac{3}{2r_0}(r-t)\right]^{1/3},\quad \tilde{H}^2_{\underline{2}} = \frac{1}{\left[\frac{3}{2}(r-t)\right]^{2/3} r_0^{1/3}},\quad \tilde{H}^3_{\underline{3}} = \frac{1}{\left[\frac{3}{2}(r-t)\right]^{2/3} r_0^{1/3}\sin\theta}. \tag{85}$$

For this metric, in (2), $\Delta\tilde{H} = 0$.

"Reduced" Hamiltonian:

$$\tilde{H}_{red} = im\gamma_{\underline{0}} - i\gamma_{\underline{0}}\left\{\tilde{H}^1_{\underline{1}}\gamma_{\underline{1}}\frac{\partial}{\partial r} + \tilde{H}^2_{\underline{2}}\gamma_{\underline{2}}\frac{\partial}{\partial\theta} + \tilde{H}^3_{\underline{3}}\gamma_{\underline{3}}\frac{\partial}{\partial\varphi}\right\}. \tag{86}$$

We insert (86) into (2) and obtain

$$\mathrm{H}_\eta = im\gamma_{\underline{0}} - i\gamma_{\underline{0}}\left\{\tilde{H}^1_{\underline{1}}\gamma_{\underline{1}}\left(\frac{\partial}{\partial r} + \frac{1}{r}\right) + \tilde{H}^2_{\underline{2}}\gamma_{\underline{2}}\left(\frac{\partial}{\partial\theta} + \frac{1}{2}\mathrm{ctg}\theta\right) + \tilde{H}^3_{\underline{3}}\gamma_{\underline{3}}\frac{\partial}{\partial\varphi}\right\} - \frac{i}{2}\gamma_{\underline{0}}\gamma_{\underline{1}}\frac{\partial\tilde{H}^1_{\underline{1}}}{\partial r}. \tag{87}$$

The Hamiltonian (87) is self-conjugate with a fairly complex time dependence.

### 5.5. Hamiltonian in the $\eta$-representation for Dirac particles in the Kruskal gravitational field

The Kruskal metric [8] is a further development of the Lemaitre-Finkelstein metric to build the most complete frame of reference for a point-mass field. The formula below, in which the frame of reference is synchronous, has been developed by I.D. Novikov [9]. In the $(\tau, R, \theta, \varphi)$ coordinates,

$$ds^2 = -d\tau^2 + \left(1+\frac{R^2}{r_0^2}\right)(1-\cos\chi)^2 dR^2 + \\ +\frac{1}{4}r_0^2\left(\frac{R^2}{r_0^2}+1\right)(1-\cos\chi)^2\left(d\theta^2+\sin^2\theta d\varphi^2\right) \quad (88)$$

$$\frac{\tau}{r_0} = \frac{1}{2}\left(\frac{R^2}{r_0^2}+1\right)^{3/2}(\pi-\chi+\sin\chi). \quad (89)$$

The determinants equal

$$g = -\left(1+\frac{R^2}{r_0^2}\right)^3(1-\cos\chi)^6\frac{1}{16}r_0^4\sin^2\theta, \\ g_G = -\left(1+\frac{R^2}{r_0^2}\right)^3(1-\cos\chi)^6\frac{1}{16}\frac{r_0^4}{R^4}. \quad (90)$$

Eqs. (88), (89) show that the metric (89) is related to the radial coordinate $R$ and proper time $\tau$ through the parameter $\eta$.

The non-zero components of the tetrad vectors in the Schwinger gauge equal

$$\tilde{H}_{\underline{0}}^0 = 1,\quad \tilde{H}_{\underline{1}}^1 = \frac{1}{\sqrt{\left(1+\frac{R^2}{r_0^2}\right)}(1-\cos\chi)},\quad \tilde{H}_{\underline{2}}^2 = \frac{2}{r_0\sqrt{\left(1+\frac{R^2}{r_0^2}\right)}(1-\cos\chi)}, \\ \tilde{H}_{\underline{3}}^3 = \frac{2}{r_0\sqrt{\left(1+\frac{R^2}{r_0^2}\right)}(1-\cos\chi)}\frac{1}{\sin\theta}. \quad (91)$$

"Reduced" Hamiltonian:

$$\tilde{H}_{red} = im\gamma_{\underline{0}} - i\gamma_{\underline{0}}\left\{\tilde{H}_{\underline{1}}^1\gamma_{\underline{1}}\frac{\partial}{\partial R} + \tilde{H}_{\underline{2}}^2\gamma_{\underline{2}}\frac{\partial}{\partial\theta} + \tilde{H}_{\underline{3}}^3\gamma_{\underline{3}}\frac{\partial}{\partial\varphi}\right\}. \quad (92)$$

According to (2), with $\Delta\tilde{H} = 0$, we have

$$\mathrm{H}_\eta = im\gamma_{\underline{0}} - i\gamma_{\underline{0}}\left\{\tilde{H}_{\underline{1}}^1\gamma_{\underline{1}}\left(\frac{\partial}{\partial R}+\frac{1}{R}\right) + \tilde{H}_{\underline{2}}^2\gamma_{\underline{2}}\left(\frac{\partial}{\partial\theta}+\frac{1}{2}\mathrm{ctg}\theta\right) + \tilde{H}_{\underline{3}}^3\gamma_{\underline{3}}\frac{\partial}{\partial\varphi}\right\} - \frac{i}{2}\gamma_{\underline{0}}\gamma_{\underline{1}}\frac{\partial\tilde{H}_{\underline{1}}^1}{\partial R}. \quad (93)$$

The derivative $\frac{\partial\tilde{H}_{\underline{1}}^1}{\partial R}$ in the last summand of (93) should allow for the dependence $\chi(R,\tau)$ (see (89)).

## 6. Axially symmetric gravitational field

### 6.1. Kerr metric in the Boyer-Lindquist coordinates

The Kerr solution in the Boyer-Lindquist coordinates [3]

$$\left(x^0, x^1, x^2, x^3\right) \equiv (t, r, \theta, \varphi) \quad (94)$$

is given by

$$g_{\alpha\beta}=\begin{array}{|c|c|c|c|}\hline -\left(1-\frac{r_0 r}{\rho^2}\right) & 0 & 0 & -\frac{a r_0 r}{\rho^2}\sin^2\theta \\ \hline 0 & \frac{\rho^2}{\Delta} & 0 & 0 \\ \hline 0 & 0 & \rho^2 & 0 \\ \hline -\frac{a r_0 r}{\rho^2}\sin^2\theta & 0 & 0 & \left(r^2+a^2+\frac{a^2 r_0 r}{\rho^2}\cdot\sin^2\theta\right)\cdot\sin^2\theta \\ \hline \end{array}\ . \qquad (95)$$

$$g=-\rho^4\cdot\sin^2\theta,\ \ g_G=-\frac{\rho^4}{r^4}\ . \qquad (96)$$

The inverse tensor has the following form:

$$g^{\alpha\beta}=\begin{array}{|c|c|c|c|}\hline -\frac{1}{\Delta}\cdot\left(r^2+a^2+\frac{a^2 r_0 r}{\rho^2}\cdot\sin^2\theta\right) & 0 & 0 & -\frac{a r_0 r}{\Delta\cdot\rho^2} \\ \hline 0 & \frac{\Delta}{\rho^2} & 0 & 0 \\ \hline 0 & 0 & \frac{1}{\rho^2} & 0 \\ \hline -\frac{a r_0 r}{\Delta\cdot\rho^2} & 0 & 0 & \frac{1}{\Delta\cdot\sin^2\theta}\left(1-\frac{r_0 r}{\rho^2}\right) \\ \hline \end{array}\ . \qquad (97)$$

Here,

$$\left.\begin{array}{l}\Delta\equiv r^2-r_0 r+a^2\\ \rho^2\equiv r^2+a^2\cdot\cos^2\theta\end{array}\right\}. \qquad (98)$$

### 6.2. Tetrad vectors in the Schwinger gauge

We will need expressions for tetrad vectors in the Schwinger gauge. The results of calculating the components of tetrad vectors $\tilde{H}^{\alpha}_{\underline{\mu}}$ are presented in Table 2. Table 3 shows the components of vectors.

Table 2.. $\tilde{H}^{\alpha}_{\underline{\mu}}$ -like tetrad vectors

| Tetrad vectors | Tetrad vector components | | | |
|---|---|---|---|---|
| $\tilde{H}^{\alpha}_{\underline{0}}$ | $\tilde{H}^{0}_{\underline{0}}=\sqrt{\left(-g^{00}\right)}$ | $\tilde{H}^{1}_{\underline{0}}=0$ | $\tilde{H}^{2}_{\underline{0}}=0$ | $\tilde{H}^{3}_{\underline{0}}=\frac{a r_0 r}{\rho^2\Delta\sqrt{\left(-g^{00}\right)}}$ |
| $\tilde{H}^{\alpha}_{\underline{1}}$ | $\tilde{H}^{0}_{\underline{1}}=0$ | $\tilde{H}^{1}_{\underline{1}}=\frac{\sqrt{\Delta}}{\rho}$ | $\tilde{H}^{2}_{\underline{1}}=0$ | $\tilde{H}^{3}_{\underline{1}}=0$ |
| $\tilde{H}^{\alpha}_{\underline{2}}$ | $\tilde{H}^{0}_{\underline{2}}=0$ | $\tilde{H}^{1}_{\underline{2}}=0$ | $\tilde{H}^{2}_{\underline{2}}=\frac{1}{\rho}$ | $\tilde{H}^{3}_{\underline{2}}=0$ |
| $\tilde{H}^{\alpha}_{\underline{3}}$ | $\tilde{H}^{0}_{\underline{3}}=0$ | $\tilde{H}^{1}_{\underline{3}}=0$ | $\tilde{H}^{2}_{\underline{3}}=0$ | $\tilde{H}^{3}_{\underline{3}}=\frac{1}{\sin\theta\cdot\sqrt{\Delta}\sqrt{\left(-g^{00}\right)}}$ |

Table 3. $\tilde{H}_{\underline{\mu}\alpha}$ -like tetrad vectors

| Tetrad vectors | Tetrad vector components | | | |
|---|---|---|---|---|
| $\tilde{H}_{\underline{0}\alpha}$ | $\tilde{H}_{\underline{0}0}=-\frac{1}{\sqrt{(-g^{00})}}$ | $\tilde{H}_{\underline{0}1}=0$ | $\tilde{H}_{\underline{0}2}=0$ | $\tilde{H}_{\underline{0}3}=0$ |
| $\tilde{H}_{\underline{1}\alpha}$ | $\tilde{H}_{\underline{1}0}=0$ | $\tilde{H}_{\underline{1}1}=\frac{\rho}{\sqrt{\Delta}}$ | $\tilde{H}_{\underline{1}2}=0$ | $\tilde{H}_{\underline{1}3}=0$ |
| $\tilde{H}_{\underline{2}\alpha}$ | $\tilde{H}_{\underline{2}0}=0$ | $\tilde{H}_{\underline{2}1}=0$ | $\tilde{H}_{\underline{2}2}=\rho$ | $\tilde{H}_{\underline{2}3}=0$ |
| $\tilde{H}_{\underline{3}\alpha}$ | $\tilde{H}_{\underline{3}0}=-\frac{ar_0 r\cdot\sin\theta}{\rho^2\sqrt{\Delta}\sqrt{(-g^{00})}}$ | $\tilde{H}_{\underline{3}1}=0$ | $\tilde{H}_{\underline{3}2}=0$ | $\tilde{H}_{\underline{3}3}=\sin\theta\cdot\sqrt{\Delta}\sqrt{(-g^{00})}$ |

**6.3. Hamiltonian $\mathrm{H}_\eta$**

First, from formula (38), we obtain

$$\begin{aligned}\tilde{H}_{red} &\equiv -\frac{im}{(-g^{00})}\tilde{\gamma}^0+\frac{i}{(-g^{00})}\tilde{\gamma}^0\tilde{\gamma}^k\frac{\partial}{\partial x^k}=\\ &=-\frac{im}{(-g^{00})}\tilde{H}^0_{\underline{0}}\gamma^{\underline{0}}+\frac{i}{(-g^{00})}\tilde{H}^0_{\underline{0}}\gamma^{\underline{0}}\left\{\tilde{H}^1_{\underline{1}}\gamma^{\underline{1}}\frac{\partial}{\partial x^1}+\tilde{H}^2_{\underline{2}}\gamma^{\underline{2}}\frac{\partial}{\partial x^2}+\tilde{H}^3_{\underline{3}}\gamma^{\underline{3}}\frac{\partial}{\partial x^3}+\tilde{H}^3_{\underline{0}}\gamma^{\underline{0}}\frac{\partial}{\partial x^3}\right\}=\\ &=\frac{im}{(-g^{00})}\tilde{H}^0_{\underline{0}}\gamma_{\underline{0}}-\frac{i}{(-g^{00})}\tilde{H}^0_{\underline{0}}\tilde{H}^1_{\underline{1}}\gamma_{\underline{0}}\gamma_{\underline{1}}\frac{\partial}{\partial r}-\\ &-\frac{i}{(-g^{00})}\tilde{H}^0_{\underline{0}}\tilde{H}^2_{\underline{2}}\gamma_{\underline{0}}\gamma_{\underline{2}}\frac{\partial}{\partial\theta}-\frac{i}{(-g^{00})}\tilde{H}^0_{\underline{0}}\tilde{H}^3_{\underline{3}}\gamma_{\underline{0}}\gamma_{\underline{3}}\frac{\partial}{\partial\varphi}-\frac{i}{(-g^{00})}\tilde{H}^0_{\underline{0}}\tilde{H}^3_{\underline{0}}\frac{\partial}{\partial\varphi}.\end{aligned} \qquad (99)$$

For the metric under consideration, $\Delta\tilde{H}$ in (2) differs from zero:

$$\Delta\tilde{H}=\frac{i}{4}\left(\frac{\partial\tilde{H}_{\underline{3}0}}{\partial r}+\frac{g^{03}}{g^{00}}\frac{\partial\tilde{H}_{\underline{3}3}}{\partial r}\right)\tilde{H}^1_{\underline{1}}S_{\underline{1}\underline{3}}+\frac{i}{4}\left(\frac{\partial\tilde{H}_{\underline{3}0}}{\partial\theta}+\frac{g^{03}}{g^{00}}\frac{\partial\tilde{H}_{\underline{3}3}}{\partial\theta}\right)\tilde{H}^2_{\underline{2}}S_{\underline{2}\underline{3}}. \qquad (100)$$

The Hamiltonian $\mathrm{H}_\eta$ is calculated using (2).

$$\begin{aligned}\mathrm{H}_\eta &=\frac{im}{(-g^{00})}\tilde{H}^0_{\underline{0}}\gamma_{\underline{0}}-\frac{i}{(-g^{00})}\tilde{H}^0_{\underline{0}}\tilde{H}^1_{\underline{1}}\gamma_{\underline{0}}\gamma_{\underline{1}}\left(\frac{\partial}{\partial r}+\frac{1}{r}\right)-\\ &-\frac{i}{(-g^{00})}\tilde{H}^0_{\underline{0}}\tilde{H}^2_{\underline{2}}\gamma_{\underline{0}}\gamma_{\underline{2}}\left(\frac{\partial}{\partial\theta}+\frac{1}{2}ctg\theta\right)-\frac{i}{(-g^{00})}\tilde{H}^0_{\underline{0}}\tilde{H}^3_{\underline{3}}\gamma_{\underline{0}}\gamma_{\underline{3}}\frac{\partial}{\partial\varphi}-\\ &-\frac{i}{2}\gamma_{\underline{0}}\gamma_{\underline{1}}\left[\frac{\partial}{\partial r}\frac{\tilde{H}^0_{\underline{0}}\tilde{H}^1_{\underline{1}}}{(-g^{00})}\right]-\frac{i}{2}\gamma_{\underline{0}}\gamma_{\underline{2}}\left[\frac{\partial}{\partial\theta}\frac{\tilde{H}^0_{\underline{0}}\tilde{H}^2_{\underline{2}}}{(-g^{00})}\right]-\frac{i}{(-g^{00})}\tilde{H}^0_{\underline{0}}\tilde{H}^3_{\underline{0}}\frac{\partial}{\partial\varphi}-\\ &-\frac{i}{4}\gamma_{\underline{3}}\gamma_{\underline{1}}\tilde{H}^1_{\underline{1}}\left(\frac{\partial\tilde{H}_{\underline{3}0}}{\partial r}+\frac{g^{03}}{g^{00}}\frac{\partial\tilde{H}_{\underline{3}3}}{\partial r}\right)+\frac{i}{4}\gamma_{\underline{2}}\gamma_{\underline{3}}\tilde{H}^2_{\underline{2}}\left(\frac{\partial\tilde{H}_{\underline{3}0}}{\partial\theta}+\frac{g^{03}}{g^{00}}\frac{\partial\tilde{H}_{\underline{3}3}}{\partial\theta}\right).\end{aligned} \qquad (101)$$

We put the tetrad vector components

$$\tilde{H}_{\underline{0}}^{0}=\sqrt{\left(-g^{00}\right)},\quad \tilde{H}_{\underline{1}}^{1}=\frac{\sqrt{\Delta}}{\rho},\quad \tilde{H}_{\underline{2}}^{2}=\frac{1}{\rho},\quad \tilde{H}_{\underline{3}}^{3}=\frac{1}{\sin\theta\cdot\sqrt{\Delta}\sqrt{\left(-g^{00}\right)}},\quad \tilde{H}_{\underline{0}}^{3}=\frac{ar_0 r}{\rho^2\Delta\sqrt{\left(-g^{00}\right)}}$$

$$\tilde{H}_{\underline{3}0}=-\frac{ar_0 r\sin\theta}{\rho^2\sqrt{\Delta}\sqrt{\left(-g^{00}\right)}},\quad \tilde{H}_{\underline{3}3}=\sin\theta\cdot\sqrt{\Delta}\sqrt{\left(-g^{00}\right)}$$

and the metric components

$$g^{00}=-\frac{1}{\Delta}\left(r^2+a^2+\frac{a^2 r_0 r}{\rho^2}\sin^2\theta\right),\quad g^{03}=-\frac{ar_0 r}{\Delta\rho^2} \tag{102}$$

into (101). Finally,

$$\begin{aligned}
\mathrm{H}_{\eta}=&\frac{im}{\sqrt{\left(-g^{00}\right)}}\gamma_{\underline{0}}-\frac{i\sqrt{\Delta}}{\rho\sqrt{\left(-g^{00}\right)}}\gamma_{\underline{0}}\gamma_{\underline{1}}\left(\frac{\partial}{\partial r}+\frac{1}{r}\right)-\\
&-\frac{i}{\rho\sqrt{\left(-g^{00}\right)}}\gamma_{\underline{0}}\gamma_{\underline{2}}\left(\frac{\partial}{\partial\theta}+\frac{1}{2}\mathrm{ctg}\theta\right)-\frac{i}{\sin\theta\cdot\left(-g^{00}\right)\sqrt{\Delta}}\gamma_{\underline{0}}\gamma_{\underline{3}}\frac{\partial}{\partial\varphi}-\frac{i}{\left(-g^{00}\right)}\frac{ar_0 r}{\rho^2\Delta}\frac{\partial}{\partial\varphi}-\\
&-\frac{i}{2}\gamma_{\underline{0}}\gamma_{\underline{1}}\left[\frac{\partial}{\partial r}\frac{\sqrt{\Delta}}{\rho\sqrt{\left(-g^{00}\right)}}\right]-\frac{i}{2}\gamma_{\underline{0}}\gamma_{\underline{2}}\left[\frac{\partial}{\partial\theta}\frac{1}{\rho\sqrt{\left(-g^{00}\right)}}\right]-\\
&-\frac{i}{4}\gamma_{\underline{3}}\gamma_{\underline{1}}\frac{\sqrt{\Delta}}{\rho}\left(-\frac{\partial}{\partial r}\frac{ar_0 r\sin\theta}{\rho^2\sqrt{\Delta}\sqrt{\left(-g^{00}\right)}}+\frac{ar_0 r}{\Delta\rho^2\left(-g^{00}\right)}\frac{\partial}{\partial r}\sin\theta\sqrt{\Delta}\sqrt{\left(-g^{00}\right)}\right)+\\
&+\frac{i}{4}\gamma_{\underline{2}}\gamma_{\underline{3}}\frac{1}{\rho}\left(-\frac{\partial}{\partial\theta}\frac{ar_0 r\sin\theta}{\rho^2\sqrt{\Delta}\sqrt{\left(-g^{00}\right)}}+\frac{ar_0 r}{\Delta\rho^2\left(-g^{00}\right)}\frac{\partial}{\partial\theta}\sin\theta\sqrt{\Delta}\sqrt{\left(-g^{00}\right)}\right).
\end{aligned} \tag{103}$$

The quantities $-g^{00},\Delta,\rho$ are defined by (102), (98).

In order to turn to the Schwarzschild Hamiltonian, one should assume that

$$a=0,\quad \Delta\to r^2-r_0 r,\quad \rho\to r,\quad \left(r^2+a^2+\frac{a^2 r_0 r}{\rho^2}\sin^2\theta\right)\to r^2. \tag{104}$$

After such a replacement, from (103), we obtain the Hamiltonian $\mathrm{H}_{\eta}$ (55) for the Schwarzschild field:

$$\mathrm{H}_{\eta}=im\sqrt{f}\gamma_{\underline{0}}-i\sqrt{f}\gamma_{\underline{0}}\left\{\gamma_{\underline{1}}\sqrt{f}\left(\frac{\partial}{\partial r}+\frac{1}{r}\right)+\gamma_{\underline{2}}\frac{1}{r}\left(\frac{\partial}{\partial\theta}+\frac{1}{2}\mathrm{ctg}\theta\right)+\gamma_{\underline{3}}\frac{1}{r\cdot\sin\theta}\frac{\partial}{\partial\varphi}\right\}-\frac{i}{2}\frac{\partial f}{\partial r}\cdot\gamma_{\underline{0}}\gamma_{\underline{1}}. \tag{105}$$

If in the expression for $\mathrm{H}_{\eta}$ (103) we restrict ourselves to the terms not higher than of the first order of smallness in the parameters $\frac{r_0}{r},\frac{ar_0}{r^2}$, we will obtain a self-conjugate Hamiltonian for the weak Kerr field.

$$\begin{aligned}
\mathrm{H}_{\eta}^{app}=&im\left(1-\frac{r_0}{2r}\right)\gamma_{\underline{0}}-i\left(1-\frac{r_0}{r}\right)\gamma_{\underline{0}}\gamma_{\underline{1}}\left(\frac{\partial}{\partial r}+\frac{1}{r}\right)-\\
&-i\left(1-\frac{r_0}{2r}\right)\gamma_{\underline{0}}\left\{\gamma_{\underline{2}}\frac{1}{r}\left(\frac{\partial}{\partial\theta}+\frac{1}{2}\mathrm{ctg}\theta\right)+\gamma_{\underline{3}}\frac{1}{r\sin\theta}\frac{\partial}{\partial\varphi}\right\}-\frac{ir_0}{2r^2}\gamma_{\underline{0}}\gamma_{\underline{1}}-\\
&-i\frac{ar_0}{r^3}\frac{\partial}{\partial\varphi}-i\frac{3}{4}\gamma_{\underline{3}}\gamma_{\underline{1}}\frac{ar_0}{r^3}\sin\theta.
\end{aligned} \tag{106}$$

In Sect. 7, $\mathrm{H}_{\eta}^{app}$ is found by the general algorithm [1], and using (1).

Previously, a self-conjugate Hamiltonian for a weak Kerr field has been obtained in Refs. [16], [19] for the metric written in isotropic coordinates.

## 7. Weak axially symmetric gravitational field

### 7.1 Kerr metric in the Boyer-Lindquist coordinates

For our purposes, we write Eqs. (95) – (98), leaving the terms not exceeding the first order of smallness in the quantities $\frac{r_0}{r}$ and $\frac{r_0 a}{r^2}$. In this approximation,

$$\Delta \approx r^2\left(1-\frac{r_0}{r}\right); \quad \rho \approx r; \quad \left(-g^{00}\right) \approx \left(1+\frac{r_0}{r}\right). \tag{107}$$

$$g_{\alpha\beta} \approx \begin{pmatrix} -\left(1-\frac{r_0}{r}\right) & 0 & 0 & -\frac{ar_0}{r}\sin^2\theta \\ 0 & \left(1+\frac{r_0}{r}\right) & 0 & 0 \\ 0 & 0 & r^2 & 0 \\ -\frac{ar_0}{r}\sin^2\theta & 0 & 0 & r^2\cdot\sin^2\theta \end{pmatrix}. \tag{108}$$

$$g^{\alpha\beta} = \begin{pmatrix} -\left(1+\frac{r_0}{r}\right) & 0 & 0 & -\frac{ar_0}{r^3} \\ 0 & \left(1-\frac{r_0}{r}\right) & 0 & 0 \\ 0 & 0 & \frac{1}{r^2} & 0 \\ -\frac{ar_0}{r^3} & 0 & 0 & \frac{1}{r^2\cdot\sin^2\theta} \end{pmatrix}. \tag{109}$$

$$g = -r^4\cdot\sin^2\theta; \quad g_G = -1; \quad \eta = \left(g_G g^{00}\right)^{1/4} = \left(1+\frac{r_0}{r}\right)^{1/4}. \tag{110}$$

### 7.2. Tetrad vectors in the Schwinger gauge

We will need expressions for tetrad vectors in the Schwinger gauge. The results of calculating the components of the tetrad vectors $\tilde{H}_{\underline{\mu}}^{\alpha}$ are presented in Table 4. Table 5 shows the components of the vectors $\tilde{H}_{\underline{\mu}\alpha}$.

Table 4. Tetrad vectors $\tilde{H}_{\underline{\mu}}^{\alpha}$

| Tetrad vectors | Tetrad vector components | | | |
|---|---|---|---|---|
| $\tilde{H}_{\underline{0}}^{\alpha}$ | $\tilde{H}_{\underline{0}}^{0} \approx \left(1+\frac{r_0}{2r}\right)$ | $\tilde{H}_{\underline{0}}^{1} = 0$ | $\tilde{H}_{\underline{0}}^{2} = 0$ | $\tilde{H}_{\underline{0}}^{3} \approx \frac{ar_0}{r^3}$ |
| $\tilde{H}_{\underline{1}}^{\alpha}$ | $\tilde{H}_{\underline{1}}^{0} = 0$ | $\tilde{H}_{\underline{1}}^{1} \approx \left(1-\frac{r_0}{2r}\right)$ | $\tilde{H}_{\underline{1}}^{2} = 0$ | $\tilde{H}_{\underline{1}}^{3} = 0$ |
| $\tilde{H}_{\underline{2}}^{\alpha}$ | $\tilde{H}_{\underline{2}}^{0} = 0$ | $\tilde{H}_{\underline{2}}^{1} = 0$ | $\tilde{H}_{\underline{2}}^{2} \approx \frac{1}{r}$ | $\tilde{H}_{\underline{2}}^{3} = 0$ |
| $\tilde{H}_{\underline{3}}^{\alpha}$ | $\tilde{H}_{\underline{3}}^{0} = 0$ | $\tilde{H}_{\underline{3}}^{1} = 0$ | $\tilde{H}_{\underline{3}}^{2} = 0$ | $\tilde{H}_{\underline{3}}^{3} \approx \frac{1}{r\cdot\sin\theta}$ |

Table 5. Tetrad vectors $\tilde{H}_{\underline{\mu}\alpha}$

| Tetrad vectors | Tetrad vector components | | | |
|---|---|---|---|---|
| $\tilde{H}_{\underline{0}\alpha}$ | $\tilde{H}_{\underline{0}0} \approx -\left(1-\frac{r_0}{2r}\right)$ | $\tilde{H}_{\underline{0}1} = 0$ | $\tilde{H}_{\underline{0}2} = 0$ | $\tilde{H}_{\underline{0}3} = 0$ |
| $\tilde{H}_{\underline{1}\alpha}$ | $\tilde{H}_{\underline{1}0} = 0$ | $\tilde{H}_{\underline{1}1} \approx \left(1+\frac{r_0}{2r}\right)$ | $\tilde{H}_{\underline{1}2} = 0$ | $\tilde{H}_{\underline{1}3} = 0$ |
| $\tilde{H}_{\underline{2}\alpha}$ | $\tilde{H}_{\underline{2}0} = 0$ | $\tilde{H}_{\underline{2}1} = 0$ | $\tilde{H}_{\underline{2}2} = r$ | $\tilde{H}_{\underline{2}3} = 0$ |
| $\tilde{H}_{\underline{3}\alpha}$ | $\tilde{H}_{\underline{3}0} \approx -\frac{ar_0}{r^2}\cdot\sin\theta$ | $\tilde{H}_{\underline{3}1} = 0$ | $\tilde{H}_{\underline{3}2} = 0$ | $\tilde{H}_{\underline{3}3} \approx r\cdot\sin\theta$ |

**7.3. Christoffel symbols**

Christoffel symbols:

$$
\begin{bmatrix}
\begin{pmatrix}0\\00\end{pmatrix}=0\\
\begin{pmatrix}0\\01\end{pmatrix}=\frac{1}{2}\frac{r_0}{r^2}\\
\begin{pmatrix}0\\02\end{pmatrix}=0\\
\begin{pmatrix}0\\03\end{pmatrix}=0\\
\begin{pmatrix}0\\11\end{pmatrix}=0\\
\begin{pmatrix}0\\12\end{pmatrix}=0\\
\begin{pmatrix}0\\13\end{pmatrix}=-\frac{3}{2}\frac{ar_0}{r^2}\sin^2\theta\\
\begin{pmatrix}0\\22\end{pmatrix}=0\\
\begin{pmatrix}0\\23\end{pmatrix}=0\\
\begin{pmatrix}0\\33\end{pmatrix}=0
\end{bmatrix}
\begin{bmatrix}
\begin{pmatrix}1\\00\end{pmatrix}=\frac{1}{2}\frac{r_0}{r^2}\\
\begin{pmatrix}1\\01\end{pmatrix}=0\\
\begin{pmatrix}1\\02\end{pmatrix}=0\\
\begin{pmatrix}1\\03\end{pmatrix}=-\frac{ar_0}{2r^2}\sin^2\theta\\
\begin{pmatrix}1\\11\end{pmatrix}=-\frac{1}{2}\frac{r_0}{r^2}\\
\begin{pmatrix}1\\12\end{pmatrix}=0\\
\begin{pmatrix}1\\13\end{pmatrix}=0\\
\begin{pmatrix}1\\22\end{pmatrix}=-r\left(1-\frac{r_0}{r}\right)\\
\begin{pmatrix}1\\23\end{pmatrix}=0\\
\begin{pmatrix}1\\33\end{pmatrix}=-r\sin^2\theta\left(1-\frac{r_0}{r}\right)
\end{bmatrix}
\begin{bmatrix}
\begin{pmatrix}2\\00\end{pmatrix}=0\\
\begin{pmatrix}2\\01\end{pmatrix}=0\\
\begin{pmatrix}2\\02\end{pmatrix}=0\\
\begin{pmatrix}2\\03\end{pmatrix}=\frac{ar_0}{r^3}\sin\theta\cos\theta\\
\begin{pmatrix}2\\11\end{pmatrix}=0\\
\begin{pmatrix}2\\12\end{pmatrix}=\frac{1}{r}\\
\begin{pmatrix}2\\13\end{pmatrix}=0\\
\begin{pmatrix}2\\22\end{pmatrix}=0\\
\begin{pmatrix}2\\23\end{pmatrix}=0\\
\begin{pmatrix}2\\33\end{pmatrix}=-\sin\theta\cos\theta
\end{bmatrix}
\begin{bmatrix}
\begin{pmatrix}3\\00\end{pmatrix}=0\\
\begin{pmatrix}3\\01\end{pmatrix}=\frac{ar_0}{2r^4}\\
\begin{pmatrix}3\\02\end{pmatrix}=-\frac{ar_0}{r^3}\frac{\cos\theta}{\sin\theta}\\
\begin{pmatrix}3\\03\end{pmatrix}=0\\
\begin{pmatrix}3\\11\end{pmatrix}=0\\
\begin{pmatrix}3\\12\end{pmatrix}=0\\
\begin{pmatrix}3\\13\end{pmatrix}=\frac{1}{r}\\
\begin{pmatrix}3\\22\end{pmatrix}=0\\
\begin{pmatrix}3\\23\end{pmatrix}=\frac{\cos\theta}{\sin\theta}\\
\begin{pmatrix}3\\33\end{pmatrix}=0
\end{bmatrix}
\qquad (111)
$$

### 7.4. Bispinor connectivities

Bispinor connectivities are calculated by the formula

$$\tilde{\Phi}_{\alpha} = \frac{1}{4}\tilde{H}^{\varepsilon}_{\underline{\mu}}\tilde{H}_{\underline{\nu}\varepsilon;\alpha}S^{\underline{\mu\nu}}. \tag{112}$$

We obtain:

$$\left.\begin{aligned}
\tilde{\Phi}_0 &= \frac{1}{4}\frac{r_0}{r^2}S_{\underline{01}} + \frac{1}{2}\frac{ar_0}{r^3}\cos\theta\cdot S_{\underline{23}} + \frac{1}{4}\frac{ar_0}{r^3}\sin\theta\cdot S_{\underline{31}}\\
\tilde{\Phi}_1 &= -\frac{3}{4}\frac{ar_0}{r^3}\sin\theta\cdot S_{\underline{03}}\\
\tilde{\Phi}_2 &= -\frac{1}{2}\left(1-\frac{r_0}{2r}\right)S_{\underline{12}}\\
\tilde{\Phi}_3 &= -\frac{3}{4}\frac{ar_0}{r^2}\sin^2\theta\cdot S_{\underline{01}} - \frac{1}{2}\cos\theta\cdot S_{\underline{23}} + \frac{1}{2}\left(1-\frac{r_0}{2r}\right)\sin\theta\cdot S_{\underline{31}}
\end{aligned}\right\}. \tag{113}$$

### 7.5. Hamiltonian $\mathrm{H}_{\eta}$

Taking into account Subsects. 7.1 – 7.4, we derive an expression for $\tilde{H}$ using (22):

$$\begin{aligned}
\tilde{\mathrm{H}} = {} & im\left(1-\frac{r_0}{2r}\right)\cdot\gamma_{\underline{0}} - i\left(1-\frac{r_0}{r}\right)\cdot\gamma_{\underline{0}}\gamma_{\underline{1}}\left(\frac{\partial}{\partial r}+\frac{1}{r}\right) -\\
& -i\left(1-\frac{r_0}{2r}\right)\frac{1}{r}\cdot\gamma_{\underline{0}}\gamma_{\underline{2}}\left(\frac{\partial}{\partial\theta}+\frac{1}{2}\mathrm{ctg}\theta\right) - i\cdot\frac{ar_0}{r^3}\frac{\partial}{\partial\varphi} -\\
& -i\left(1-\frac{r_0}{2r}\right)\frac{1}{r\sin\theta}\cdot\gamma_{\underline{0}}\gamma_{\underline{3}}\frac{\partial}{\partial\varphi} - i\cdot\gamma_{\underline{0}}\gamma_{\underline{1}}\frac{r_0}{4r^2} - i\frac{3}{4}\frac{ar_0}{r^3}\sin\theta\cdot S_{\underline{31}}.
\end{aligned} \tag{114}$$

Since the Kerr solution is stationary, the general formula for $\mathrm{H}_{\eta}$,

$$\mathrm{H}_{\eta} = \tilde{\eta}\tilde{H}\tilde{\eta}^{-1} + i\tilde{\eta}\frac{\partial\tilde{\eta}^{-1}}{\partial t}, \tag{115}$$

in our case will be written as

$$\mathrm{H}_{\eta} = \tilde{\eta}\tilde{H}\tilde{\eta}^{-1}, \tag{116}$$

where $\eta$ is defined by (110).

As a result, the Hamiltonian $\mathrm{H}_{\eta}$ can be written as

$$\begin{aligned}
\mathrm{H}_{\eta} = {} & im\left(1-\frac{r_0}{2r}\right)\cdot\gamma_{\underline{0}} - i\left(1-\frac{r_0}{r}\right)\cdot\gamma_{\underline{0}}\gamma_{\underline{1}}\left(\frac{\partial}{\partial r}+\frac{1}{r}\right) -\\
& -i\left(1-\frac{r_0}{2r}\right)\frac{1}{r}\cdot\gamma_{\underline{0}}\gamma_{\underline{2}}\left(\frac{\partial}{\partial\theta}+\frac{1}{2}\mathrm{ctg}\theta\right) - i\gamma_{\underline{0}}\gamma_{\underline{1}}\frac{r_0}{2r^2} -\\
& -i\gamma_{\underline{3}}\gamma_{\underline{1}}\frac{3}{4}\frac{ar_0}{r^3}\sin\theta - i\left(1-\frac{r_0}{2r}\right)\frac{1}{r\sin\theta}\cdot\gamma_{\underline{0}}\gamma_{\underline{3}}\frac{\partial}{\partial\varphi} - i\cdot\frac{ar_0}{r^3}\frac{\partial}{\partial\varphi}.
\end{aligned} \tag{117}$$

Eq. (117) coincides with Eq. (106), derived by expanding the general expression for $\mathrm{H}_{\eta}$ (103). In turn, the general expression (103) is obtained using (2).

Analogously, using (114), we can easily check if the formula (1) is valid for the metric under consideration (108).

Thus, the same expression for $\mathrm{H}_{\eta}$ for a weak Kerr field is in fact derived in three different ways.

For the block-diagonal metrics like (3), as exemplified by the Kerr metric with the formula (2), we can see that the algorithm for finding the Dirac self-conjugate Hamiltonians with a flat scalar product becomes significantly simpler.

**8. Open Friedmann model**

Consider the case of the open Friedmann model in the coordinates

$$\left(x^0, x^1, x^2, x^3\right) = \left(t, \chi, \theta, \varphi\right).$$

For this model, the non-stationary metric takes the following form:

$$ds^2 = -dt^2 + a^2(t)\left(d\chi^2 + \mathrm{sh}^2\chi\left[d\theta^2 + \sin^2\theta d\varphi^2\right]\right), \tag{118}$$

$$g = -a^6 \mathrm{sh}^4\chi \sin^2\theta, \quad g_G = -a^6. \tag{119}$$

The non-zero components of the tetrad vectors $\tilde{H}^{\alpha}_{\underline{\alpha}}$ in the Schwinger gauge equal

$$\left.\begin{array}{l} \left\{\tilde{H}^0_{\underline{0}} = 1; \tilde{H}^1_{\underline{1}} = \dfrac{1}{a}; \tilde{H}^2_{\underline{2}} = \dfrac{1}{a\cdot \mathrm{sh}\chi}; \tilde{H}^3_{\underline{3}} = \dfrac{1}{a\cdot \mathrm{sh}\chi\cdot \sin\theta};\right. \\ \left\{\tilde{H}_{\underline{00}} = -1; \tilde{H}_{\underline{11}} = a; \tilde{H}_{\underline{22}} = a\cdot \mathrm{sh}\chi; \tilde{H}_{\underline{33}} = a\cdot \mathrm{sh}\chi\cdot\sin\theta;\right. \\ \left\{\tilde{H}^{\underline{0}}_{\alpha} = 1; \tilde{H}^{\underline{1}}_1 = a; \tilde{H}^{\underline{2}}_2 = a\cdot \mathrm{sh}\chi; \tilde{H}^{\underline{3}}_3 = a\cdot \mathrm{sh}\chi\cdot\sin\theta;\right. \\ \left\{\tilde{H}^{\underline{00}} = -1; \tilde{H}^{\underline{11}} = \dfrac{1}{a}; \tilde{H}^{\underline{22}} = \dfrac{1}{a\cdot \mathrm{sh}\chi}; \tilde{H}^{\underline{33}} = \dfrac{1}{a\cdot \mathrm{sh}\chi\cdot\sin\theta}\right. \end{array}\right\}. \tag{120}$$

Calculations of the Hamiltonian $\tilde{H}$ give

$$\tilde{H} = im\gamma_{\underline{0}} - i\gamma_{\underline{0}}\frac{1}{a}\left\{\gamma_{\underline{1}}\left(\frac{\partial}{\partial\chi} + \mathrm{cth}\chi\right) + \gamma_{\underline{2}}\frac{1}{\mathrm{sh}\chi}\left(\frac{\partial}{\partial\theta} + \frac{1}{2}\mathrm{ctg}\theta\right) + \right.$$
$$\left. + \gamma_{\underline{3}}\frac{1}{\mathrm{sh}\chi\cdot\sin\theta}\frac{\partial}{\partial\varphi}\right\} - i\frac{3}{2}\frac{\dot{a}}{a}. \tag{121}$$

The Hamiltonian $\mathrm{H}_\eta$ is defined by (2). For this metric, $\Delta\tilde{H} = 0$

$$H_\eta = im\gamma_{\underline{0}} - i\gamma_{\underline{0}}\frac{1}{a}\left\{\gamma_{\underline{1}}\left(\frac{\partial}{\partial\chi} + \mathrm{cth}\chi\right) + \gamma_{\underline{2}}\frac{1}{\mathrm{sh}\chi}\left(\frac{\partial}{\partial\theta} + \frac{1}{2}\mathrm{ctg}\theta\right) + \right.$$
$$\left. + \gamma_{\underline{3}}\frac{1}{\mathrm{sh}\chi\cdot\sin\theta}\frac{\partial}{\partial\varphi}\right\}. \tag{122}$$

In the quasi-stationary approximation, for the cosmological time $t$, the energy operator for a particle moving in the $\chi$-direction equals

$$E = \sqrt{\mathrm{H}^2_\eta} = \sqrt{m^2 + \frac{\mathbf{p}^2_\chi}{a^2(t)}}. \tag{123}$$

Here, $\mathbf{p}_\chi = -i\left(\dfrac{\partial}{\partial\chi} + \mathrm{cth}\chi\right)$.

Let us denote

$$a(t)\mathrm{sh}\chi = \frac{a(t)}{a_0}a_0\mathrm{sh}\chi = b(t)a_0\mathrm{sh}\chi = b(t)r, \tag{124}$$

where $b(t_0) = 1$; zero subscripts correspond to the current time $\left(t \le t_0\right)$.

If the radius of the spatial curvature of the universe currently goes to infinity, $\left(a_0 \to \infty\right)$, then

$$r \approx a_0\chi. \tag{125}$$

In this case, for the spatially flat Friedmann model, the Hamiltonian (122) becomes equal to

$$\mathrm{H}_\eta = im\gamma_{\underline{0}} - i\gamma_{\underline{0}}\gamma_{\underline{1}}\frac{1}{b(t)}\left(\frac{\partial}{\partial r} + \frac{1}{r}\right) - i\gamma_{\underline{0}}\gamma_{\underline{2}}\frac{1}{b(t)r}\left(\frac{\partial}{\partial\theta} + \frac{1}{2}\mathrm{ctg}\theta\right) - i\gamma_{\underline{0}}\gamma_{\underline{3}}\frac{1}{b(t)r\sin\theta}\frac{\partial}{\partial\varphi}. \tag{126}$$

In Cartesian coordinates, the expression for $\mathrm{H}_\eta$ is

$$\mathrm{H}_{\eta} = im\gamma_{\underline{0}} - \frac{i}{b(t)}\gamma_{\underline{0}}\gamma_{\underline{k}}\frac{\partial}{\partial x^{k}}. \tag{127}$$

**9. Clifford torus metric**

The metric proposed in [10] in the $(t,\varphi_1,\varphi_2,z)$ coordinates is given by

$$\begin{aligned} ds^2 &= -dt^2 + \left(\rho_1(z)\right)^2 d\varphi_1^2 + \left(\rho_2(z)\right)^2 d\varphi_2^2 + \left[1+\rho_1'(z)^2 + \rho_2'(z)^2\right]dz^2, \\ g &= g_G = -\rho_1^2\rho_2^2\left(1+\rho_1'^2+\rho_2'^2\right). \end{aligned} \tag{128}$$

In (128), the prime denotes the derivative with respect to the $z$ coordinate.
Tetrad vectors in the Schwinger gauge:

$$\begin{aligned} &\tilde{H}^0_{\underline{0}} = 1, \quad \tilde{H}^1_{\underline{0}} = 0, \quad \tilde{H}^2_{\underline{0}} = 0, \quad \tilde{H}^3_{\underline{0}} = 0; \\ &\tilde{H}^0_{\underline{1}} = 0, \quad \tilde{H}^1_{\underline{1}} = \frac{1}{\rho_1}, \quad \tilde{H}^2_{\underline{1}} = 0, \quad \tilde{H}^3_{\underline{1}} = 0; \\ &\tilde{H}^0_{\underline{2}} = 0, \quad \tilde{H}^1_{\underline{2}} = 0, \quad \tilde{H}^2_{\underline{2}} = \frac{1}{\rho_2}, \quad \tilde{H}^3_{\underline{2}} = 0; \\ &\tilde{H}^0_{\underline{3}} = 0, \quad \tilde{H}^1_{\underline{3}} = 0, \quad \tilde{H}^2_{\underline{3}} = 0, \quad \tilde{H}^3_{\underline{3}} = \frac{1}{\rho_3}. \end{aligned} \tag{129}$$

In (129), $\rho_3(z) = \sqrt{1+\left(\rho'(z)\right)^2 + \left(\rho_2'(z)\right)^2}$. In accordance with (38), the "reduced" Hamiltonian $\tilde{H}_{red}$ equals

$$\tilde{H}_{red} = im\gamma_{\underline{0}} - \frac{i}{\rho_1}\gamma_{\underline{0}}\gamma_{\underline{1}}\frac{\partial}{\partial\varphi_1} - \frac{i}{\rho_2}\gamma_{\underline{0}}\gamma_{\underline{2}}\frac{\partial}{\partial\varphi_2} - \frac{i}{\rho_3}\gamma_{\underline{0}}\gamma_{\underline{3}}\frac{\partial}{\partial z}. \tag{130}$$

For this metric, in (2), $\Delta\tilde{H} = 0$. Then, the Hamiltonian $H_\eta$ in accordance with (2) equals

$$\begin{aligned} H_\eta &= \frac{1}{2}\left(\tilde{H}_{red} + \tilde{H}^{+}_{red}\right) = \\ &= im\gamma_{\underline{0}} - \frac{i}{\rho_1}\gamma_{\underline{0}}\gamma_{\underline{1}}\frac{\partial}{\partial\varphi_1} - \frac{i}{\rho_2}\gamma_{\underline{0}}\gamma_{\underline{2}}\frac{\partial}{\partial\varphi_2} - \frac{i}{\rho_3}\gamma_{\underline{0}}\gamma_{\underline{3}}\frac{\partial}{\partial z} + \frac{i}{2}\gamma_{\underline{0}}\gamma_{\underline{3}}\frac{\rho_3'}{\rho_3^2}. \end{aligned} \tag{131}$$

**10. Equivalence of Hamiltonians with harmonic Cartesian or Boyer-Lindquist coordinates in a weak Kerr field**

As we know, harmonic coordinates satisfy the condition formulated by Th. De-Donder and V.A.Fock [12], [13].

In Refs. [16], [19], the following form of self-conjugate Dirac Hamiltonian $H_c$ was derived using harmonic Cartesian coordinates for a weak Kerr field:

$$\begin{aligned} H_c &= im\left(1-\frac{r_0}{2r}\right)\gamma_{\underline{0}} - i\left(1-\frac{r_0}{r}\right)\gamma_{\underline{0}}\gamma_{\underline{k}}\frac{\partial}{\partial x^k} - \\ &-i\frac{r_0}{2r^3}\gamma_{\underline{0}}\gamma_{\underline{k}}x_k - i\frac{r_0 a}{r^3}\left(x_1\frac{\partial}{\partial x_2} - x_2\frac{\partial}{\partial x_1}\right) + \\ &+i\frac{r_0 a}{4r^3}\left[\gamma_{\underline{1}}\gamma_{\underline{2}}\left(1-3\frac{x_3^2}{r^2}\right) - \gamma_{\underline{2}}\gamma_{\underline{3}}\frac{3x_3x_1}{r^2} - \gamma_{\underline{3}}\gamma_{\underline{1}}\frac{3x_3x_2}{r^2}\right]. \end{aligned} \tag{132}$$

Similar to Subsects 6, 7, $\mathbf{a} = (0,0,a)$; $\mathbf{J} = \frac{r_0}{2}\mathbf{a}$ in (132) is angular momentum of a rotating source of the Kerr field.

When using Boyer-Lindquist coordinates [3], the self-conjugate Hamiltonian $H_{B-L}$ for a weak Kerr field is defined by (106), (117):

$$H_{B-L}=im\left(1-\frac{r_0}{2r}\right)\gamma_{\underline{0}}-i\left(1-\frac{r_0}{r}\right)\gamma_{\underline{0}}\gamma_{\underline{1}}\left(\frac{\partial}{\partial r}+\frac{1}{r}\right)-$$
$$-i\left(1-\frac{r_0}{2r}\right)\frac{1}{r}\left[\gamma_{\underline{0}}\gamma_{\underline{2}}\left(\frac{\partial}{\partial\theta}+\frac{1}{2}\operatorname{ctg}\theta\right)+\gamma_{\underline{0}}\gamma_{\underline{3}}\frac{1}{\sin\theta}\frac{\partial}{\partial\varphi}\right]- \tag{133}$$
$$-\gamma_{\underline{0}}\gamma_{\underline{1}}\frac{r_0}{2r^2}-i\frac{r_0a}{r^3}\frac{\partial}{\partial\varphi}-i\frac{3}{4}\frac{r_0a}{r^3}\gamma_{\underline{3}}\gamma_{\underline{1}}\sin\theta.$$

In (132), (133), the summands without $a$ correspond to the Schwarzschild metric. These parts of (132), (133) are physically equivalent to each other.

The last but one summands in (132), (133) are also physically equivalent to each other. Indeed, in a weak Kerr field, Boyer-Lindquist coordinates are reduced to spherical coordinates.

Hence,

$$x_3=r\cdot\cos\theta,$$
$$x_2=r\cdot\sin\theta\sin\varphi, \tag{134}$$
$$x_1=r\cdot\sin\theta\cos\varphi.$$

$$\frac{\partial}{\partial\varphi}=\frac{\partial}{\partial x_2}\frac{\partial x_2}{\partial\varphi}+\frac{\partial}{\partial x_1}\frac{\partial x_1}{\partial\varphi}=x_1\frac{\partial}{\partial x_2}-x_2\frac{\partial}{\partial x_1}. \tag{135}$$

Given (135), we can see the desired equivalence.

As for the last summands in (132), (133), they do not seem to be equivalent at first. Suffice it to note that the last summand in (132) contains three spin matrices $\left(i\gamma_{\underline{1}}\gamma_{\underline{2}}=\Sigma_3, i\gamma_{\underline{2}}\gamma_{\underline{3}}=\Sigma_1, i\gamma_{\underline{3}}\gamma_{\underline{1}}=\Sigma_2\right)$, while the last summand in (133) has only one spin matrix $\Sigma_2$. This may put the correctness of (133) into question, although it was derived in Subsect. 7 in three different ways.

In order to resolve this, let us write the Hamiltonian (133) using the representation of Dirac matrices in spherical coordinates (79). The matrices (79) are related to the matrices $\gamma_{\underline{1}},\gamma_{\underline{2}},\gamma_{\underline{3}}$ by the unitary transformation (80).

The Hamiltonian (133) with the local Dirac matrices $\gamma_{\underline{r}},\gamma_{\underline{\theta}},\gamma_{\underline{\varphi}}$ takes the following form:

$$H'_{B-L}=im\left(1-\frac{r_0}{2r}\right)\gamma_{\underline{0}}-i\left(1-\frac{r_0}{r}\right)\gamma_{\underline{0}}\gamma_{\underline{r}}\frac{\partial}{\partial r}-$$
$$-i\left(1-\frac{r_0}{2r}\right)\gamma_{\underline{0}}\gamma_{\underline{\theta}}\frac{1}{r}\frac{\partial}{\partial\theta}-i\left(1-\frac{r_0}{2r}\right)\gamma_{\underline{0}}\gamma_{\underline{\varphi}}\frac{1}{r\cdot\sin\theta}\frac{\partial}{\partial\varphi}- \tag{136}$$
$$-\gamma_{\underline{0}}\gamma_{\underline{r}}\frac{r_0}{2r^2}-i\frac{r_0a}{r^3}\frac{\partial}{\partial\varphi}+i\frac{1}{4}\frac{r_0a}{r^3}\Big[\gamma_{\underline{1}}\gamma_{\underline{2}}\left(1-3\cos^2\theta\right)-$$
$$-3\gamma_{\underline{2}}\gamma_{\underline{3}}\cos\theta\sin\theta\cos\varphi-3\gamma_{\underline{3}}\gamma_{\underline{1}}\cos\theta\sin\theta\sin\varphi\Big].$$

Given (134), the last summands in (136) and (132) coincide with each other. The Hamiltonian (136) is physically equivalent to the Hamiltonian (133), since it is obtained using the unitary transformation (80):

$$H'_{B-L}=RH_{B-L}R^{+}. \tag{137}$$

The analysis indirectly proves that Eq. (103) is valid for the general Hamiltonian in Boyer-Lindquist coordinates. Eq. (103) can be used for Kerr gravitational field of arbitrary strength and angular momentum of the field source of arbitrary magnitude.

The results obtained above demonstrate that for clear physical interpretation of individual summands of Dirac Hamiltonians one should use harmonic Cartesian coordinates. Classical interpretation of individual Hamiltonian terms requires transition to the Foldy-Wouthuysen representation [20], [21].

**Conclusions**

This study develops the algorithm proposed in [1] for constructing self-conjugate Hamiltonians $\mathrm{H}_\eta$ in the $\eta$-representation with a flat scalar product to describe the dynamics of Dirac particles in arbitrary gravitational fields. We prove that a Hamiltonian in the $\eta$-representation for any gravitational field, including a time-dependent field, is a Hermitian part of the initial Dirac Hamiltonian $\tilde{H}$ derived using tetrad vectors in the Schwinger gauge. We also prove that for the block-diagonal matrices like (3), the Hamiltonian $\mathrm{H}_\eta$ can be calculated by the formula (2) using "reduced" parts of the Hamiltonians $\tilde{H}$ and $\tilde{H}^+$ without or with a small number of summands with bispinor connectivities. Using this method, we for the first time find self-conjugate Hamiltonians $\mathrm{H}_\eta$ for the Kerr metric in the Boyer-Lindquist form and for the Eddington-Finkelstein, Finkelstein-Lemaitre, Kruskal, Clifford torus metrics and also for non-stationary metrics of open and spatially flat Friedmann models.

In this paper, we also prove physical equivalence of Dirac Hamiltonians in a weak Kerr field in harmonic Cartesian and Boyer-Lindquist coordinates. We point at the necessity of using harmonic Cartesian coordinates for clear physical interpretation of individual terms in the Hamiltonians.

In [22], the algorithm for deriving self-conjugate Dirac Hamiltonians in the $\eta$-representation is extended to the electromagnetic case. The Hamiltonian derived is applied to the case when the nonstationary gravitational field describes the spatially flat Friedmann model, and the electromagnetic field is an extension of the Coulomb potential to the case of this model.

Following other authors [17], we demonstrate that energy levels in atomic systems are invariable in cosmological time.

Spectral lines of atoms in the spatially flat Friedmann model are identical at different points of cosmological time, and redshift is attributed completely to the growth of the wavelength of photons in the expanding universe.

At the same time, we observed that interaction forces and physical dimensions of atomic and quark bound systems vary with universe expansion.

The expressions for Hamiltonians $\mathrm{H}_\eta$, derived in this paper, can also be employed to study the behavior of Dirac particles in the vicinity of black holes, and scattering and absorption of such particles by black holes.

**Acknowledgement**

The authors would like to thank Prof. P.Fiziev for the useful discussions, advices and criticism.